\documentclass[aps,pra,twocolumn,showpacs,10pt]{revtex4-1}

\usepackage{braket}
\usepackage{amsmath}
\usepackage{color}
\usepackage{graphicx}
\usepackage{subcaption}
\usepackage{caption}
\usepackage{amssymb}
\usepackage{amsfonts}
\usepackage{tikz}
\usepackage{hyperref}
\usepackage[format=plain,justification=centerlast,singlelinecheck=true]{caption}

\begin{document}





\title{Degradation of entanglement between two accelerated parties:\\ Bell states under the Unruh effect}


\author{Benedikt Richter$^{1,2}$  }
\author{Yasser Omar$^{1,2,3}$}
\affiliation{$^1$Physics of Information Group, Instituto de Telecommunica\c{c}\~oes, Portugal}
\affiliation{$^2$Instituto Superior T\'{e}cnico, Universidade de Lisboa, Portugal}
\affiliation{$^3$CEMAPRE, ISEG, Universidade de Lisboa, Portugal}
\date{14 August 2015}

\begin{abstract}

We study the entanglement of families of Unruh modes in the Bell states $|\Phi^\pm\rangle =1/\sqrt{2}(|00\rangle\pm|11\rangle)$ and $|\Psi^\pm\rangle=1/\sqrt{2}(|01\rangle\pm|10\rangle)$ shared by two accelerated observers and find fundamental differences in the robustness of entanglement against acceleration for these states. States $\Psi^\pm$ are entangled for all finite accelerations, whereas, due to the Unruh effect, states $\Phi^\pm$ lose their entanglement for finite accelerations. This is true for Bell states of two bosonic modes, as well as for Bell states of a bosonic and a fermionic mode. Furthermore, there are also differences in the degradation of entanglement for Bell states of fermionic modes. We reveal the origin of these distinct characteristics of entanglement degradation and discuss the role that is played by particle statistics. Our studies suggest that the behavior of entanglement in accelerated frames strongly depends on the occupation patterns of the constituent states, whose superposition constitutes the entangled state, where especially states $\Phi^\pm$ and $\Psi^\pm$  exhibit distinct characteristics regarding entanglement degradation. Finally, we point out possible implications of hovering over a black hole for these states.

\end{abstract}

\pacs{03.67.Mn,  03.65.Ud,  04.62.+v \hspace{13mm} Journal reference: 	Phys. Rev. A \textbf{92}, 022334 (2015)}
\maketitle
\section{Introduction}

Entanglement and quantum correlations in general play an important role in different areas of physics, as, for example, in quantum information \cite{nielsen2010quantum} and black holes \cite{braunstein2013better, raey}.  Furthermore, it is known that the quantum correlations of an entangled state shared by accelerated observers are not invariant with respect to acceleration but are altered by the Unruh effect \cite{fuentes2005alice}. Interestingly, the noninvariance of quantum correlations in this relativistic regime can be employed to carry out quantum information tasks \cite{bruschi2013relativistic, martin2013processing, friis2012motion, bradler2009private, friis2013relativistic}. Although accelerated motion can, in some special cases, create entanglement between Unruh modes \cite{montero2011entangling},  generally entanglement is degraded due to the Unruh effect.

In the past, the degradation of entanglement in bipartite states composed of Unruh modes shared by an inertial observer and an uniformly accelerated one was studied in detail \cite{alsing2006entanglement, pan2008degradation, martin2009fermionic, bruschi2010unruh, martin2011redistribution, pan2008hawking, martin2010quantum1,  bruschi2012particle, bruschi2010unruh, martin2010unveiling, leon2009spin, pan2008degradation,  montero2011entanglement, martin2010population, chang2012entanglement}. For fermionic fields, entanglement approaches a finite value in the infinite acceleration limit \cite{alsing2006entanglement}, while for bosons it vanishes asymptotically  \cite{fuentes2005alice}. One way to study the entanglement between two accelerated observers is to analyze an entangled state shared by three parties, where two parties are in accelerated motion, and subsequently trace out the inertial observer \cite{ shamirzaie2012tripartite, khan2014non, wang2011multipartite}. A more natural way to study entanglement in this framework is to restrict to bosonic entanglement in accelerated two-mode squeezed states and use tools from continuous variable quantum mechanics \cite{ahn2007hawking, adesso2007continuous}. There it was found that, for these squeezed states, entanglement vanishes for finite accelerations, in contrast to the entanglement of states shared by an accelerated observer and an inertial one. 

As realized more recently in \cite{dragan2013localized, dragan2013localized2, doukas2013entanglement}, there are some caveats in the interpretation of states of Unruh modes. Still, however, the use of Unruh modes, which allows for closed analytical solutions, provides a valuable framework to understand the mechanisms that lead to a decrease of quantum correlations in entangled states when described by accelerated observers. The goal of this work is to provide further insight into the degradation of entanglement that occurs when entangled states are observed by accelerated parties. Therefore, in this work, we study families of states composed of Unruh modes that are maximally entangled from the inertial perspective and investigate the residual entanglement when these states are seen by uniformly accelerated observers.

We start by studying the entanglement between two accelerated observers sharing the fermionic Bell states $|\Phi^\pm\rangle=1/\sqrt{2}(|0_\omega 0_\Omega\rangle\pm|1_\omega 1_\Omega\rangle)$ and $|\Psi^\pm\rangle=1/\sqrt{2}(|0_\omega 1_\Omega\rangle\pm|1_\omega 0_\Omega\rangle)$. We find that entanglement is nonvanishing for all accelerations, and that the degradation of quantum correlations depends on the specific state shared by the parties. The reason for the survival of entanglement for fermions is rooted in the statistics obeyed by fermions, while the statedependence of entanglement degradation seems to be originated in the occupation pattern of the state that is shared between the two observers. By occupation pattern we mean the pattern of both constituent states (in the following just called constituents), for example, $|00\rangle$, $|01\rangle$, $|10\rangle$, and $|11\rangle$, whose superposition defines the entangled state. That is, the set of excitations created by the Unruh effect depends on the state one starts with.

Furthermore, we study maximally entangled bosonic states shared by two accelerated observers and, in contrast to \cite{ahn2007hawking, adesso2007continuous, adesso2012continuous}, we consider two bosonic modes of frequency $\omega$ and $\Omega$, respectively, in the Bell states $\Phi^\pm$ and $\Psi^\pm$. We find that the Unruh effect degrades entanglement in these two states very differently. While, as in \cite{ahn2007hawking, adesso2007continuous, adesso2012continuous}, bosonic modes in state $\Phi^\pm$ lose all their entanglement for finite accelerations, the entanglement of these modes in state $\Psi^\pm$ is nonvanishing for all finite accelerations. We find that this crucial difference in the degradation of entanglement is due to the differing occupation patterns of the constituents of the two Bell states, and is manifest in the appearance of a ``cut-off function'' in the expression of the negativity.

Then we extend our studies to accelerated states of a bosonic mode maximally entangled with a fermionic one. Thereby we find states whose negativity factorizes. However, due to the bosonic mode involved in these states, there is no entanglement surviving in the infinite acceleration limit. Then, moving on to Bell states $\Phi^\pm$ and $\Psi^\pm$, we obtain qualitatively the same behavior as for the purely bosonic Bell states. Thus, we find evidence for the importance of the occupation patterns of the constituents. The particular occupation patterns of its constituents can protect a state's entanglement against the effects of acceleration, i.e., the Unruh effect. Using an effective state picture we are able to explain the differences in the behavior.

In the past, the different behavior of the entanglement of accelerated fermions and bosons led to some discussions \cite{martin2010population, martin2009fermionic,  bruschi2012particle}. Here we address this issue and discuss the role played by particle statistics by combining the results of the fermion-fermion, boson-boson and the boson-fermion cases. We conclude that there are essentially two factors determining the fading of entanglement, where one is purely from particle statistics and one strongly depends on the occupation patterns of the constituents of the state that is considered. Finally, we discuss some effects that hovering over a black hole at a fixed distance from the horizon has on the entanglement of the states that are studied in this work. 

The outline is the following. In Sec.\ \ref{qftrindler} we give a short introduction to quantum fields in Rindler space. In Secs.\ \ref{ffe}, \ref{bbe}, and \ref{bfe} we study the entanglement of fermion-fermion, boson-boson and boson-fermion Bell states in accelerated motion, respectively. Then, in Sec.\ \ref{secstat}, we discuss the role of particle statistics for entanglement degradation and give two factors that determine the characteristics of said degradation. In Sec.\ \ref{secbh} we outline possible implications of our findings for Bell states in the vicinity of a black hole and, finally, in Sec.\ \ref{conclusions}, we give the conclusions of this work.

For the sake of brevity, throughout this work, we call the occupation patterns of the constituents of a state just the structure of a state.

\section{Quantum fields in Rindler space} \label{qftrindler}
\begin{figure}
\center
\includegraphics[width=0.9\columnwidth]{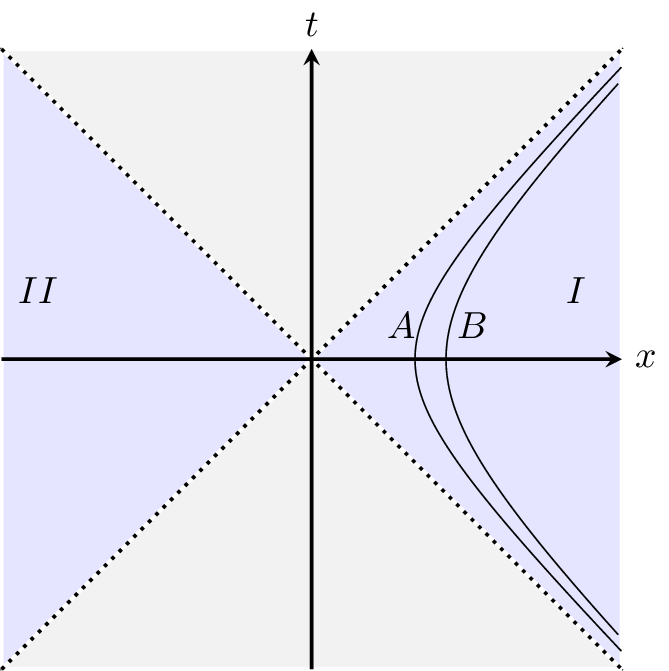}
\caption{(Color online) Rindler space: Regions $I$ and $II$ are causally disconnected due to the presence of horizons at $x=t$ and $x=-t$. The worldlines of two observers $A$ and $B$ in (differently) accelerated motion are shown. The states shared by $A$ and $B$ are prepared and distributed in the asymptotic past.} \label{figrind}
\end{figure}

 We give a brief introduction to quantum field theory in Rindler space. The main purpose is to introduce the framework we are using in this work. More details can be found, for example, in \cite{birrell1984quantum, bruschi2010unruh, takagi1986vacuum}. We work in units where $c=\hbar=k_B=1$. Minkowski  coordinates $(t,x)$ and Rindler coordinates $(\xi, \eta)$ are related by the transformations
\begin{subequations}
\begin{align}
t=& \xi \sinh(\eta),\\
x=& \xi \cosh(\eta),
\end{align}
\end{subequations}
where the range of $\xi$ and $\eta$ is given by $-\infty<\xi, \eta<\infty$. Notice that $\xi$ is positive in the right wedge (region $I$) and $\xi$ is negative in the left wedge (region $II$). Then we obtain the following metric
\begin{equation}\label{metricrindler}
ds^2=\xi^2 d\eta^2-d\xi^2.
\end{equation}
Considering a world line with $\xi(\tau)=\frac{1}{a}$, where $\tau$ is the proper time along this trajectory and $|a|$ is the proper acceleration, we find $\eta(\tau)=a\tau$. Thus, in Minkowski coordinates the world line reads $t(\tau)=\frac{1}{a} \sinh(a\tau)$, $x(\tau)=\frac{1}{a} \cosh(a\tau)$. As shown in Fig.\ \ref{figrind}, the two regions $I$ and $II$ of Rindler space are causally disconnected due to the presence of horizons at $x=t$ and $x=-t$. A timelike Killing vector in region $I$ is given by $\partial_\eta$ ($-\partial_\eta$ in $II$). 

\subsection{Bosons}\label{qftboson}

We consider the quantization of a massless scalar field $\phi$ (see \cite{birrell1984quantum, takagi1986vacuum} for details). We quantize fields with respect to the Killing vectors $\partial_\eta$ and $-\partial_\eta$ independently in the two regions. The Klein-Gordon equation $\Box \phi=0$ in Rindler coordinates has solutions which depend on $\eta$ as \cite{takagi1986vacuum}
\begin{equation}
u_{\tilde{\omega}}^\pm\propto e^{\pm i{\tilde{\omega}}\eta},\hspace{5mm} \pm\partial_\eta u_{\tilde{\omega}}^\pm=i{\tilde{\omega}} u_{\tilde{\omega}}^\pm,
\end{equation}
where ${\tilde{\omega}}$ is a positive parameter and the sign $\pm$ depends on the Rindler wedge. These modes are positive frequency modes with respect to the respective timelike Killing vectors. We denote by $u_{\tilde{\omega}}^{I,II}$, i.e., $u_{\tilde{\omega}}^{I}\propto e^{- i{\tilde{\omega}}\eta}$ and $u_{\tilde{\omega}}^{II}\propto e^{+ i{\tilde{\omega}}\eta}$, the solutions in regions $I$ and $II$, respectively. We can expand $\phi$ in this basis and obtain 
\begin{equation}
\phi=\int_0^\infty d{\tilde{\omega}}\left(a_{\tilde{\omega}}^I u_{\tilde{\omega}}^{I}+a_{\tilde{\omega}}^{I\dagger} u_{\tilde{\omega}}^{I*}+a_{\tilde{\omega}}^{II} u_{\tilde{\omega}}^{II}+a_{\tilde{\omega}}^{II\dagger} u_{\tilde{\omega}}^{II*}\right),
\end{equation}
where the $a_{\tilde{\omega}}^{I/II}$ and $a_{\tilde{\omega}}^{I/II \dagger}$ are the usual commuting annihilation and creation operators in regions $I$ and $II$, respectively. The dependence on the proper time $\tau=\frac{\eta}{a}$ is given by $u_{\tilde{\omega}}^{I}\propto e^{- ia{\tilde{\omega}}\tau}$. Therefore, the energy $\omega$ seen by the accelerated observer is given by $\omega=a{\tilde{\omega}}$. Remember that in Minkowski space $\phi$ can be expanded as
\begin{equation}
\phi=\int_{0}^\infty d\omega_M\left(a_{\omega_M}^M u_{\omega_M}^{M}+a_{\omega_M}^{M\dagger} u_{\omega_M}^{M*}\right),
\end{equation}
where $a_{\omega_M}^{M}$ and $a_{\omega_M}^{M \dagger}$ are the commuting  Minkowski annihilation and creation operators. These two expansions lead to different Fock spaces. Consider the Minkowski ($M$) vacuum $|0\rangle_M$ and the Rindler ($R$) vacuum $|0\rangle_R=|0\rangle_I\otimes|0\rangle_{II}$ that are defined as
\begin{subequations}
\begin{align}
a_{\omega_M}^M|0\rangle_M=&0,\\
a_{\tilde{\omega}}^I |0\rangle_R=&a_{\tilde{\omega}}^{II}|0\rangle_R=0,
\end{align}
\end{subequations}
and, in general, $a_{\tilde{\omega}}^{I,II } |0\rangle_M\neq 0$. Next, we want to introduce the so-called Unruh basis that we use in the following. The Rindler creation and annihilation operators $a_{\tilde{\omega}}^{I,II}$ are related to the corresponding Unruh ($U$) ones $a_{\tilde{\omega}}^{U,1,2}$ by a Bogoliubov transformation as follows \cite{birrell1984quantum}:
\begin{subequations}\label{bogo}
\begin{align}
a_{\tilde{\omega}}^{I}=&\frac{1}{\sqrt{2\sinh(\pi{\tilde{\omega}})}}\left(e^{\frac{\pi{\tilde{\omega}}}{2}} a_{\tilde{\omega}}^{U,2}+e^{-\frac{\pi{\tilde{\omega}}}{2}}a_{\tilde{\omega}}^{U,1\dagger}\right),\\
a_{\tilde{\omega}}^{II}=&\frac{1}{\sqrt{2\sinh(\pi{\tilde{\omega}})}}\left(e^{\frac{\pi{\tilde{\omega}}}{2}} a_{\tilde{\omega}}^{U,1}+e^{-\frac{\pi{\tilde{\omega}}}{2}}a_{\tilde{\omega}}^{U,2\dagger}\right).
\end{align}
\end{subequations}
These share the positive frequency analyticity properties of the $u_{\omega_M}^M$ and therefore have the same vacuum state $a_{\tilde{\omega}}^{U,1,2}|0\rangle_M= 0$, i.e., $|0\rangle_U=|0\rangle_M$ \cite{unruh1976notes}.

In this work we study accelerated observers confined to Rindler wedge $I$ and work in the Unruh basis. To obtain the appropriate description of what an accelerated observer is experiencing, we use Bogoliubov transformations (\ref{bogo}) to go from the Unruh basis to the Rindler basis. For an Unruh mode ${\tilde{\omega}}$, the vacuum and one particle states are given by 
\begin{subequations}\label{bosstates1}
\begin{align}
|0_{\tilde{\omega}}\rangle_U=& \sum_n \frac{\tanh^{n} (r)}{\cosh (r)} |n_{\tilde{\omega}}\rangle_I |n_{\tilde{\omega}}\rangle_{II},\\
|1_{\tilde{\omega}}\rangle_U=&\sum_n \frac{\tanh^{n} (r)}{\cosh^2 (r)} \sqrt{n+1}|n+1_{\tilde{\omega}}\rangle_I |n_{\tilde{\omega}}\rangle_{II},\label{bosstates1sp}
\end{align}
\end{subequations}
for a massless uncharged scalar field, where the acceleration parameter $r$ is set by  $\tilde{\omega}$ and they are related by $r=\text{arctanh} (e^{-\pi{\tilde{\omega}}})$. Further, $|n\rangle_I$ and $|n\rangle_{II}$ are the $n$-particle states in regions $I$ and $II$, respectively. Note that in (\ref{bosstates1sp}) we denote the one-particle state  $ a_{\tilde{\omega}}^{U,2\dagger}\,|0_{\tilde{\omega}}\rangle_U$ by $|1_{\tilde{\omega}}\rangle_U$. That is, the excitation is localized in $I$. Similarly, for a  massless charged scalar field the vacuum and one-particle states are given by \cite{bruschi2012particle}
\begin{subequations}\label{bosstates2}
\begin{align}
|0_{\tilde{\omega}}\rangle_U=& \sum_{n,m}\frac{\tanh^{n+m} (r)}{\cosh^2 (r)}  |n m\rangle_I |n m\rangle_{II},\\
|1_{\tilde{\omega}}\rangle_U^+=& \sum_{n,m} \frac{\tanh^{n+m} (r)}{\cosh^3 (r)} \sqrt{n+1}|(n+1) m\rangle_I |n m\rangle_{II},\\
|1_{\tilde{\omega}}\rangle_U^-=&\sum_{n,m} \frac{\tanh^{n+m} (r)}{\cosh^3 (r)} \sqrt{m+1}|n (m+1)\rangle_I |n m\rangle_{II},
\end{align}
\end{subequations}
where $|n m\rangle_{R}$ denotes the state of $m/ n$ particles/ anti-particles of energy ${\tilde{\omega}}$ in region $R=I, II$.

\subsection{Fermions}

We model the fermionic field by a massless Grassmannian valued scalar field $\psi$. Then the quantization of  $\psi$ can be carried out analogously to the bosonic case.  To obtain the appropriate description of what an accelerated observer is experiencing, we use the Bogoliubov transformations  \cite{bruschi2010unruh}
\begin{subequations}\label{bogof}
\begin{align}
a_{\tilde{\omega}}^{I}=&\frac{1}{\sqrt{2\cosh(\pi{\tilde{\omega}})}}\left(e^{\frac{\pi{\tilde{\omega}}}{2}} a_{\tilde{\omega}}^{U,2}+e^{-\frac{\pi{\tilde{\omega}}}{2}}a_{\tilde{\omega}}^{U,1\dagger}\right),\\
a_{\tilde{\omega}}^{II}=&\frac{1}{\sqrt{2\cosh(\pi{\tilde{\omega}})}}\left(e^{\frac{\pi{\tilde{\omega}}}{2}} a_{\tilde{\omega}}^{U,1}+e^{-\frac{\pi{\tilde{\omega}}}{2}}a_{\tilde{\omega}}^{U,2\dagger}\right)
\end{align}
\end{subequations}
 to go from the Unruh basis to the Rindler basis.  We choose the notation $| ijkl \rangle_{\tilde{\omega}}=| i_{\tilde{\omega}} \rangle_I^+ \otimes | j_{\tilde{\omega}} \rangle_{II}^- \otimes | k_{\tilde{\omega}} \rangle_I^- \otimes | l_{\tilde{\omega}}\rangle_{II}^+$, where $+/-$ denote particles and anti-particles, respectively. We use the chosen order throughout the work and obtain for an Unruh mode  ${\tilde{\omega}}$   
\begin{subequations}\label{fermiinrindler}
\begin{align}
|0_{\tilde{\omega}}^F\rangle_U=& \cos^2 (r_f)| 0000 \rangle_{\tilde{\omega}}-\cos (r_f)\sin (r_f)| 0011 \rangle_{\tilde{\omega}}\nonumber\\
+&\cos (r_f)\sin (r_f)| 1100 \rangle_{\tilde{\omega}}-\sin^2 (r_f)| 1111 \rangle_{\tilde{\omega}},\label{fermiinrindlervac}\\
|1_{\tilde{\omega}}^F\rangle_U^+=& \cos (r_f)| 1000 \rangle_{\tilde{\omega}}-\sin (r_f)| 1011 \rangle_{\tilde{\omega}},\\
|1_{\tilde{\omega}}^F\rangle_U^-=& \cos (r_f)| 0010 \rangle_{\tilde{\omega}}+\sin (r_f)| 1110 \rangle_{\tilde{\omega}},
\end{align}
\end{subequations}
where the acceleration parameter $r_f$ is given by $r_f=\arctan (e^{-\pi {\tilde{\omega}}})$. 

\subsection{Unruh effect}
An accelerated observer does not necessarily agree with an inertial observer on the number of particles in a given state. Consider, for example, the Minkowski vacuum state $|0\rangle_M=|0\rangle_U\equiv|0\rangle$. Then the vacuum expectation value of the (Rindler) number operator $\langle0|a_{\tilde{\omega}}^{I/II\dagger}a_{\tilde{\omega}}^{I/II}|0\rangle$ can be calculated using (\ref{bogof}), leading to
\begin{equation}\label{unruhvacex}
\langle0|a_{\tilde{\omega}}^{I/II\dagger}a_{\tilde{\omega}}^{I/II}|0\rangle=1+e^{\frac{2\pi\omega}{a}}.
\end{equation}
Thus, we see that an accelerated observer perceives the Minkowski vacuum as a thermal state of temperature $T_U$ (Unruh temperature) \cite{unruh1976notes},
\begin{equation}\label{unruhtemp}
T_U=\frac{a}{2\pi}.
\end{equation}
Further, we introduce the fermionic and the bosonic partition functions $Z_F^\omega$ and $Z_B^\omega$ that are given by
\begin{align}
Z_B^\omega=&\frac{1}{1-e^{-\frac{2\pi\omega}{a}} }=\frac{1}{1-e^{-\frac{\omega}{T_U}}}, \label{partbos}\\
Z_F^\omega=&1+e^{-\frac{2\pi\omega}{a}}=1+e^{-\frac{\omega}{T_U}}. \label{partfer}
\end{align}

There are some subtleties when working with the global Unruh modes  (\ref{bosstates1}), (\ref{bosstates2}), and (\ref{fermiinrindler}), as pointed out in \cite{dragan2013localized, dragan2013localized2, doukas2013entanglement}. Therefore, one point that we want to emphasize is  the implicit dependence on the acceleration $a$ in (\ref{bogo}) and (\ref{bogof}) through the relation ${\tilde{\omega}}= \frac{\omega}{a}$. As a consequence, after fixing the frequency $\omega$, each of the Unruh modes (\ref{bosstates1}), (\ref{bosstates2}), and (\ref{fermiinrindler}) forms a family of modes that is labeled by $a$. That is, by varying the acceleration parameter $r/r_f$ (or equivalently $a$), one also varies the particular Unruh mode under consideration. In order to pick a particular state, $\omega$ and $a$ have to be fixed. Intuitively, one can say that the acceleration $a$ is already encoded in the Unruh modes.  We revisit these issues when we discuss the entanglement of fermions in Sec. \ref{ffe}. In the following, for the sake of simplicity, we omit the tilde in $\tilde{\omega}$ whenever it is clear from the context whether we are talking about $\omega$ or $\tilde{\omega}$.

In this brief introduction to quantum fields in Rindler space we have set up the tools and notation we are using in the following. In the next section we study the degradation of entanglement in fermionic Bell states due to acceleration.

\section{Entanglement and entropy of uniformly accelerated fermion states} \label{ffe}
Due to the anticommutativity of fermionic creation and annihilation operators (Pauli exclusion principle), there is only a finite number of maximally entangled states of two modes of a  fermionic field. Considering particle states, there are just two possible maximally entangled states. These are the two Bell states of two fermionic modes ($FF$):
\begin{subequations}\label{statesfermi}
\begin{align}
|\Psi_{FF}^\pm\rangle=&\frac{1}{\sqrt{2}}\left(|1^F_\omega\rangle_U^+|0^F_\Omega\rangle_U\pm|0^F_\omega\rangle_U|1^F_\Omega\rangle_U^+\right),\\
|\Phi_{FF}^\pm\rangle=&\frac{1}{\sqrt{2}}\left(|0^F_\omega\rangle_U|0^F_\Omega\rangle_U\pm|1^F_\omega\rangle_U^+|1^F_\Omega\rangle_U^+\right).
\end{align}
\end{subequations}
 The subscript $U$ emphasizes that we are working in the Unruh basis.

\subsection{Negativity}\label{entangff}
We consider two families of entangled fermionic Unruh modes $\omega$ and $\Omega$ undergoing constant accelerations $a_\omega$ and $a_\Omega$, respectively. The acceleration parameters of the modes are denoted by $r_f^\omega$ and $r_f^\Omega$. Therefore, starting from the families of states $\{\psi_i\}=\{\Psi_{FF}^\pm, \Phi_{FF}^\pm\}$ written in the Unruh basis, we use (\ref{fermiinrindler}) to obtain the density matrices $\rho_{I,II}^{(i)}$ 
\begin{equation}\label{densitymat}
\rho_{I,II}^{(i)}=|\psi_i\rangle\langle\psi_i|.
\end{equation}
To describe the system as it is seen by an observer confined to region $I$, we have to trace out modes that have their support in the inaccessible region. Then the reduced density matrix $\rho_i$ is given by
\begin{equation}\label{reddensitymat}
\rho_i=Tr_{II} \left(\rho_{I,II}^{(i)}\right).
\end{equation}
As a measure of entanglement we use the negativity $N$ (defined in Appendix \ref{appneg}), which is an entanglement monotone \cite{vidal2002computable}. It is known that a bipartite state is not separable if its negativity is nonzero \cite{peres1996separability}. Furthermore, the vanishing of the negativity provides a necessary and sufficient condition for the separability of mixed states of two qubits \cite{horodecki1996separability}. Fermions, in general, cannot be treated as qubits. However,  when charge superselection is respected, two fermionic modes can be represented as two qubits \cite{friis2013fermionic}. A further property of the negativity is that a state with vanishing negativity contains no distillable (free) entanglement, although, in this case, there can be nondistillable (bound) entanglement present \cite{horodecki1998mixed}. In this work, we ignore the possibility of bound entanglement and refer to free entanglement as entanglement.

To obtain the negativities $N_i$ of states $\psi_i$, we have to calculate the partially transposed reduced density matrices $\rho_i^{pT}$. We find that these matrices are block diagonal. More details of the calculations can be found in  Appendix \ref{appneg}. The final results for the negativities $N_i$ read
\begin{align}
N_{\Psi_{FF}^\pm}=&\frac{1}{2}\left(\sqrt{\frac{1}{Z_F^\omega}\frac{1}{Z_F^\Omega}+\left(\frac{n_F^\omega+n_F^\Omega}{2}\right)^2}-\frac{n_F^\omega+ n_F^\Omega}{2}\right),\label{negastate2}\\
N_{\Phi_{FF}^\pm}=&\frac{1}{2}\,\frac{1}{Z_F^\omega}\frac{1}{Z_F^\Omega},\label{negastate1}
\end{align}
where  $T_{\omega/\Omega}$ are the Unruh temperatures (\ref{unruhtemp}) corresponding to the respective accelerations $a_\omega$ and $a_\Omega$, $Z_F^{\omega/\Omega}$ is the partition function (\ref{partfer}), and $\omega$, $\Omega$  are the energies of the modes. Further, we introduced the occupation numbers $n_F^{\omega}=(1+ e^{\omega/T_{\omega}})^{-1}$ and $n_F^{\Omega}=(1+ e^{\Omega/T_{\Omega}})^{-1}$.

Having obtained the analytic expressions for the negativities of the families of  maximally entangled fermion states (\ref{statesfermi}), we want to comment on the physical interpretation of these states. As discussed in Sec.\ \ref{qftrindler}, we are not describing a fixed state $\psi_i$, but rather describe a two-parameter family of states $\psi_i$ labeled by $a_\omega$ and $a_\Omega$. Therefore, the negativities (\ref{negastate2}) and (\ref{negastate1}) give the entanglement of the states $\psi_i$ (maximally entangled from the inertial perspective), when the two modes $\omega$ and $\Omega$ are seen by accelerated observers undergoing the accelerations $a_\omega$ and $a_\Omega$, respectively. Equivalently, we can think of states $\psi_i$ as families labeled by $\omega$ and $\Omega$, when we are fixing $a_\omega$ and $a_\Omega$. It should be noted that for a given set $(\omega, \Omega, a_\omega, a_\Omega)$ the only difference between these states is the difference in their occupation pattern, in the sense of $|00\rangle+|11\rangle$ vs. $|10\rangle+|01\rangle$. In the following, we discuss the effects of acceleration on these families of states and, for the sake of brevity, refer to them just as states.

\begin{figure}
\includegraphics[width=\columnwidth]{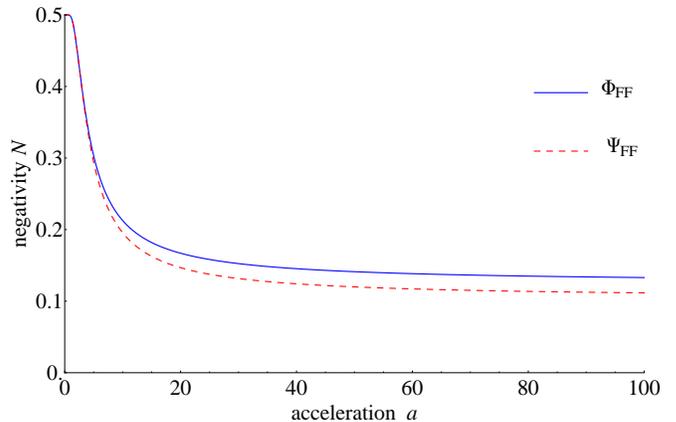}
\caption{\label{picff}(Color online) Negativities for maximally entangled fermion states $\{\psi_i\}$ versus the acceleration  $a=a_\omega=a_\Omega$, measured in units of  $\frac{1}{L}$ (for some length scale $L$), for frequencies $\omega=\Omega=\frac{1}{L}$. For each fixed acceleration $a$, the entanglement degradation for state $\Phi_{FF}^\pm$ (blue continuous line) is stronger than for state $\Psi_{FF}^\pm$ (red dashed line). The finite asymptotic values for states $\Phi_{FF}^\pm$ and $\Psi_{FF}^\pm$ are $1/8$ and $1/4(\sqrt{2}-1)$, respectively. }
\end{figure}

Considering state $\Phi_{FF}^\pm$, it is interesting to note that the negativity (\ref{negastate1}) factorizes as 
\begin{equation}\label{prodstru}
N_{\Phi_{FF}^\pm}(r_f^\omega, r_f^\Omega)= 2 N_{f}(r_f^\omega)N_{f}(r_f^\Omega),
\end{equation}
where we denoted $N_{\Phi_{FF}^\pm}(r_f^\omega, r_f^\Omega =0)$ by $N_{f}(r_f^\omega)$. Note that $N_{f}(r_f^\omega)$ is the negativity in case of only one mode being seen by an accelerated observer (acceleration parameter $r_f^\omega$). The negativity $N_{f}(r_f^\omega)=\frac{1}{2}\cos ^2(r_f^\omega)$ was obtained, for example, in \cite{martin2009fermionic}. This product structure is absent for state $\Psi_{FF}^\pm$, where the negativity is given by (\ref{negastate2}). Thus, the degradation of entanglement in the case of a fermionic field shows no universal behavior. Indeed different classes of states, $\Psi_{FF}^\pm$ and $\Phi_{FF}^\pm$, are not equally robust against acceleration; see Fig. \ref{picff}.  That is a feature that was absent in previous studies of one accelerated observer.

There is a fundamental difference between states $\Phi_{FF}^\pm$ and $\Psi_{FF}^\pm$. While each state $\psi_i$ is a superposition of two states (constituents) of the form $|k^F_\omega\rangle_U|l^F_\Omega\rangle_U$ ($k, l\in\{0,1^+\}$), only for state $\Psi_{FF}^\pm$ both such states lead to a contribution to a (the same) diagonal element of the reduced density matrix that is relevant for entanglement. More precisely, for non-vanishing acceleration,  $|1^F_\omega\rangle_U^+|0^F_\Omega\rangle_U$ as well as $|0^F_\omega\rangle_U|1^F_\Omega\rangle_U^+$ contribute to the matrix element $|1^F_\omega\rangle^+_I |1^F_\Omega\rangle^+_I  \langle 1^F_\omega|_I^+  \langle 1^F_\Omega|_I^+$ of the reduced density matrix. This is well reflected in the expression for the negativity, given by (\ref{negastate2}). The contribution of $|1^F_\omega\rangle_U^+|0^F_\Omega\rangle_U$ is quantified by $n_F^\Omega$ and the one of $|0^F_\omega\rangle_U|1^F_\Omega\rangle_U^+$ by $n_F^\omega$.  Due to the symmetry, (\ref{negastate2}) depends only the average occupation number $\bar{n}_F=1/2(n_F^\omega+n_F^\Omega)$. This behavior distinguishes $\Psi_{FF}^\pm$ from $\Phi_{FF}^\pm$.

 As in the setting of one mode seen by an accelerated observer and one mode seen by an inertial observer \cite{bruschi2010unruh, martin2009fermionic, leon2009spin}, in the limit of infinite acceleration the negativity does not vanish, but it approaches a finite limit. The surviving entanglement is calculated to be
\begin{subequations}\label{limitff}
\begin{align}
\lim_{r_f^\omega,r_f^\Omega\to\infty} N_{\Phi_{FF}^\pm}=&\frac{1}{8},\\
\lim_{r_f^\omega,r_f^\Omega\to\infty} N_{\Psi_{FF}^\pm}=&\frac{1}{4}\left(\sqrt{2}-1\right).
\end{align}
\end{subequations}
The fact that there is entanglement surviving in this limit is specific for initially pure maximally entangled states and contrasts with the case of starting from a tripartite state, where one observer is inertial and two observers are accelerated. In that case, after tracing out the inertial observer, the bipartite entanglement between the modes observed by accelerated observers vanishes in this limit \cite{shamirzaie2012tripartite}. Numerical studies of fermionic mixed state entanglement also showed that  entanglement is extinguished for most states in the infinite acceleration limit \cite{montero2011fermionic2}.

Our results reduce to the known results for one accelerated observer if one takes the limit $r_f^\Omega\to 0$ and we obtain the universal behavior reported in \cite{martin2009fermionic},
\begin{equation}\label{negoneobs}
\lim_{r_f^\Omega\to0} N_{i} \equiv N_f= \frac{1}{2}\cos ^2(r_f^\omega)
\end{equation}
and thus
\begin{equation}
\lim_{r_f^\omega\to\infty} N_f=\frac{1}{4}.
\end{equation}
 Interestingly, the behavior of the negativity under acceleration does not depend on whether there is entanglement created in some sectors. We define a sector of a state $\psi_i$ as follows: A sector of state $\psi_i$ consists of all the elements of the reduced density matrix $\rho_i$ that contribute to one block of the block diagonal partially transposed reduced density matrix $\rho_i^{pT}$. For example, $\Phi^\pm_{FF}$ has four sectors, as can be seen from the partially transposed reduced density matrix $\rho_{\Phi^\pm_{FF}}^{pT}$ (see  Appendix \ref{appneg}). When the acceleration is increasing from zero, entanglement decreases in the sector where it is initiated and, depending on the particular structure of the state, entanglement is created in previously nonentangled sectors. More details can be found in Appendix \ref{appa2}.
 
The consequences of the fact that states (\ref{fermiinrindler}) depend on the acceleration $a_{\omega/\Omega}$ only via the ratios $\omega/a_{\omega}$ and $\Omega/a_{\Omega}$ are manifest in (\ref{negastate2}) and (\ref{negastate1}). One observes that high-frequency modes are less effected by acceleration than low-frequency modes are. This is due to the larger wavelength of low-energy modes. The larger the wavelength, and therefore the spatial extension, compared to the inverse Unruh temperature, the more the system gets ``stretched'' by the acceleration. Thus, the effects of acceleration are stronger in this case and the rate of entanglement degradation is higher.
  


After this detailed study of the negativity of states (\ref{statesfermi}), we now move on to analyze the entropy and the mutual information of these states. This provides further insight into the effects acceleration has on the correlations in fermion states.

\subsection{Entropy and mutual information}\label{entropyff}

\begin{figure}[t]
\center
\includegraphics[width=\columnwidth]{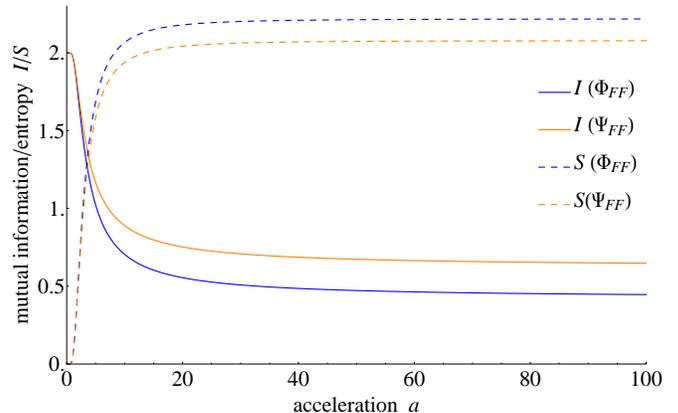}
\caption{(Color online) Mutual information (dashed lines), measured in bits, and von Neumann entropy (continuous lines) for maximally entangled fermion states $\{\psi_i\}$ plotted  versus the acceleration  $a=a_\omega=a_\Omega$, measured in units of  $\frac{1}{L}$ (for some length scale $L$), for frequencies $\omega=\Omega=\frac{1}{L}$.  The mutual information of states $\Phi_{FF}^\pm$ and $\Psi_{FF}^\pm$, as well as the entropies of these states, approach different asymptotic values.}\label{mutinfo}
\end{figure}

In the following, we analyze entropy and mutual information of the fermion states. Since we are considering accelerated observers and therefore trace out region $II$ to obtain the reduced density matrices $\rho_i$, the resulting state is not pure any more. Thus, the entropy of $\rho_i$ increases due to entanglement with modes in region $II$. Among the different measures of entropy, the most widely used is the  von Neumann entropy $S$, given by
\begin{equation}\label{vonneu}
S(\rho_i)=-Tr_I \left(\rho_{i}\ln(\rho_{i})\right).
\end{equation}
In our setting, the von Neumann entropy can be calculated analytically. Since the corresponding expressions are quite long and not very enlightening, we give the plots of $S$ as a function of the acceleration in Fig.\ \ref{mutinfo}. In the limit of infinite acceleration the von Neumann entropies approach the asymptotic values $S^\infty(\rho_i)$, which are given in the following
\begin{subequations}
\begin{align}
S^\infty(\rho_{\Phi_{FF}^\pm})=&  \frac{\ln (32)}{4}-\frac{3+2 \sqrt{2}}{8}  \ln \left(\frac{3+2 \sqrt{2}}{32} \right) \nonumber \\
+&\frac{2 \sqrt{2}-3}{8}  \ln \left(\frac{3-2 \sqrt{2}}{32} \right),  \\
S^\infty(\rho_{\Psi_{FF}^\pm})=& \ln\left(8\right).
\end{align}
\end{subequations}

Furthermore, we calculate the mutual information $I$ between modes $\omega$ and $\Omega$ as a measure of quantum and classical correlations as
\begin{equation}\label{mutinofmration}
I_i=S\left(Tr_\omega\left(\rho_i\right)\right)+S\left(Tr_\Omega\left(\rho_i\right)\right)-S(\rho_i),
\end{equation}
where $Tr_{\omega/\Omega}$ denotes the trace over mode $\omega/\Omega$. The resulting mutual information of states $\{\psi_i\}$ (in bits) is shown in Fig.\ \ref{mutinfo}. In the limit of infinite acceleration the surviving correlations are given by $I_i^\infty$:
\begin{subequations}\label{limitmut}
\begin{align}
I_{\Phi_{FF}^\pm}^\infty=&  \frac{1}{8\ln\left(2\right)} (-\ln \left(\frac{531441}{256}\right)+\nonumber\\
+&\left(3+2 \sqrt{2}\right) \ln \left(3+2 \sqrt{2}\right) +\nonumber\\
+&\left(3-2 \sqrt{2}\right) \ln \left(3-2 \sqrt{2}\right)), \\
I_{\Psi_{FF}^\pm}^\infty=&  \frac{1}{\ln\left(2\right)} \ln\left(\frac{8}{3\sqrt{3}}\right).
\end{align}
\end{subequations}

As can be seen from Fig.\ \ref{mutinfo}, the entanglement entropies vanish for zero acceleration, as the mode is localized in region $I$. So there is no entanglement between modes in region $I$ and modes in region $II$. As the acceleration increases, an acceleration horizon forms and the entanglement entropy increases due to tracing out modes with support in the region behind the horizon. We see that the entanglement between modes in the accessible region and modes in the inaccessible region does not increase equally for all states, but depends on the particular $\psi_i$. The mutual information of states $\Phi_{FF}^\pm$  and $\Psi_{FF}^\pm$ decreases with increasing acceleration and, in the infinite acceleration limit, approaches distinct values ($I_{\Phi_{FF}^\pm}^\infty\approx 0.4$, $I_{\Psi_{FF}^\pm}^\infty\approx 0.6$). Since $I_i^\infty<1$ for all states, we conclude that also classical correlations become degraded with increasing acceleration. 

In this section we studied the degradation of quantum and classical correlations in fermion states that is caused by uniform acceleration. In the following section we study the entanglement in bosonic Bell states.

\section{Entanglement of uniformly accelerated boson-boson states}\label{bbe}

We continue by investigating the entanglement of Bell states of Unruh modes $\omega$ and  $\Omega$ of a massless uncharged scalar field,
\begin{subequations}\label{statesbb}
\begin{align}
|\Psi^\pm_{BB}\rangle=&\frac{1}{\sqrt{2}}\left(|0_\omega\rangle_U|1_\Omega\rangle_U\pm|1_\omega\rangle_U|0_\Omega\rangle_U\right),\\
|\Phi^\pm_{BB}\rangle=&\frac{1}{\sqrt{2}}\left(|0_\omega\rangle_U|0_\Omega\rangle_U\pm|1_\omega\rangle_U|1_\Omega\rangle_U\right),
\end{align}
\end{subequations}
where $\omega$, $\Omega$ are the frequencies and $0$, $1$ the occupation numbers of the Unruh modes. We consider the two modes $\omega$ and $\Omega$ undergoing constant accelerations $a_\omega$ and $a_\Omega$, respectively. The acceleration parameters of the modes are denoted by $r^\omega$ and $r^\Omega$. We write states (\ref{statesbb}) in the Rindler basis to obtain the infinite-dimensional density matrices $\rho_{I,II}^{(i=\Psi^\pm,\Phi^\pm)}$. Then, to describe the system as it is seen by an observer confined to region $I$, we have to trace out modes that have their support in region $II$.

\begin{figure}[]
\center
\includegraphics[width=\columnwidth]{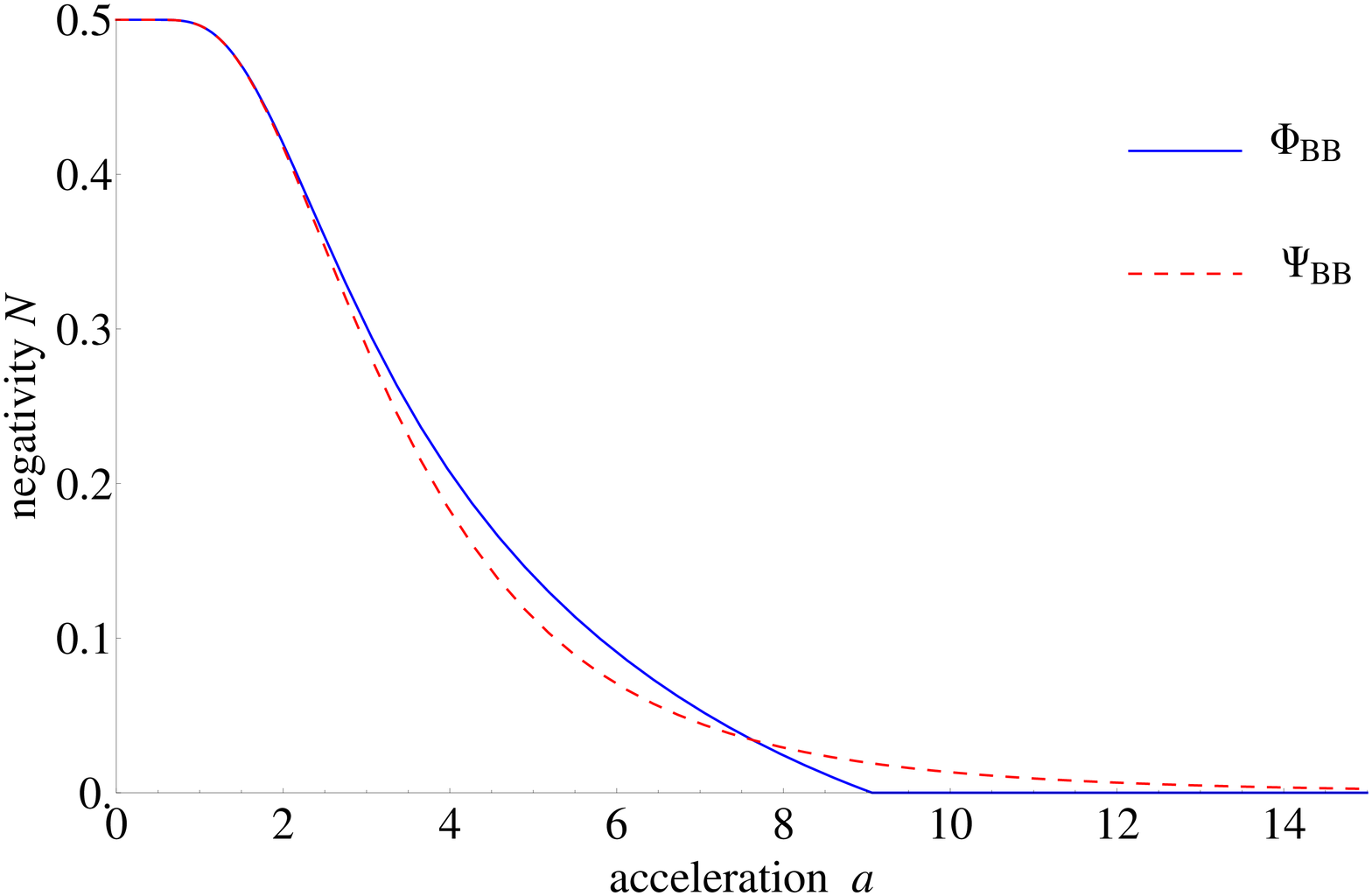}
\caption{(Color online) Negativities for the maximally entangled boson states $\Phi^\pm_{BB}$ and $\Psi^\pm_{BB}$, where both observers are accelerated,  plotted   versus the acceleration  $a=a_\omega=a_\Omega$, measured in units of  $\frac{1}{L}$ (for some length scale $L$), for frequencies $\omega=\Omega=\frac{1}{L}$. The negativity of state $\Phi^\pm_{BB}$ (blue continuous line) vanishes for finite accelerations, while the negativity of $\Psi^\pm_{BB}$ (red dashed line) vanishes asymptotically.}
\label{picbb}
\end{figure}

 As in Sec.\ \ref{ffe}, to obtain the negativities $N_i$ of states (\ref{statesbb}), we determine the partially transposed reduced density matrices $\rho_i^{pT}$ that are block diagonal and calculate the negative eigenvalues. More details can be found in Appendix \ref{appbb}. The negativities of  Bell states (\ref{statesbb}) are given by the expressions
\begin{widetext}
\begin{align}
N_{\Psi^\pm_{BB}}=&\frac{1}{2}\frac{1}{\left(Z_B^\omega\right)^2}\frac{1}{\left(Z_B^\Omega\right)^2}\left(\sqrt{Z_B^\omega Z_B^\Omega+\frac{1}{4}\left(n_B^\omega+n_B^\Omega\right)^2}-\frac{1}{2}\left(n_B^\omega+n_B^\Omega\right)\right)+\sum_{n=1}^\infty N_{\Psi^\pm_{BB}}^{(n)},\label{N002}\\
N_{\Phi^\pm_{BB}}=& \frac{1}{2}\frac{1}{\left(Z_B^\omega\right)^2}\frac{1}{\left(Z_B^\Omega\right)^2}\, \gamma_{\Phi^\pm_{BB}}\left(n_B^\omega,\, n_B^\Omega\right)+\sum_{n=1}^\infty N_{\Phi^\pm_{BB}}^{(n,0)}+\sum_{m=1}^\infty N_{\Phi^\pm_{BB}}^{(0,m)}, \label{N0011}
\end{align}
\end{widetext}
where $n_B^{\omega}=(e^{\frac{\omega}{T_{\omega}}}-1)^{-1}$ [$n_B^{\Omega}=(e^{\frac{\Omega}{T_{\Omega}}}-1)^{-1}$]  is the Bose-Einstein distribution with the Unruh temperatures $T_{\omega/\Omega}=\frac{a_{\omega/\Omega}}{2\pi}$, $Z_B^{\omega/\Omega}$ is the bosonic partition function (\ref{partbos}), and $\gamma_{\Phi^\pm_{BB}}$ is  given by 
\begin{equation}
\gamma_{\Phi^\pm_{BB}}=1-n_B^\omega  n_B^\Omega.
\end{equation}
$N_{\Psi^\pm_{BB}}^{(n)}$, $N_{\Phi^\pm_{BB}}^{(n,0)}$, and $N_{\Phi^\pm_{BB}}^{(0,m)}$ give small corrections compared to the leading term and can be found in Appendix \ref{appbb}. The degradation of entanglement shows fundamentally different characteristics for the two Bell states $\Psi^\pm_{BB}$ and $\Phi^\pm_{BB}$; see Fig.\ \ref{picbb}. While for $\Psi^\pm_{BB}$ entanglement vanishes asymptotically, $\Phi^\pm_{BB}$ loses all its entanglement for finite acceleration.

In case of $\Psi^\pm_{BB}$, for $r^\omega=r^\Omega$, only one of the blocks on the diagonal of the partially transposed reduced density matrix admits negative eigenvalues. There is no entanglement generated in any sector, and only the sector where  entanglement is initialized  contributes to the negativity $N_{\Psi^\pm_{BB}}$. From now on, we refer to a sector as all elements of the reduced density matrix that contribute to one block of the block diagonal partially transposed reduced density matrix. For  $r^\omega\neq r^\Omega$ there is entanglement created in all sectors. 

Note that the negativity vanishes asymptotically. The reason why $\Psi^\pm_{BB}$ does not become nondistillable (and therefore does not become separable)  for any finite acceleration is that the occupation of state $ |0 0 \rangle$ is always zero; i.e., the Unruh effect does not drive the occupation of this state. This is due to the fact that the constituents of $\Psi^\pm_{BB}$ both contain one excitation and, thus, state $ |0 0 \rangle$ is not accessible. In consequence the matrix element $ |0 0 \rangle\langle 0 0 |$ in $\rho_{\Psi^\pm_{BB}}$ is always zero and entanglement vanishes only  asymptotically for infinite acceleration, as in this regime all occupation is shifted towards highly excited states.

As we show in Appendix \ref{appbb}, the negativity of $\Phi^\pm_{BB}$ is of the form $N_{\Phi^\pm_{BB}}=\sum_{n, m=0}^\infty N_{\Phi^\pm_{BB}}^{(n, m)}$ and it can be seen that each of the $N_{\Phi^\pm_{BB}}^{(n, m)}$ is bounded from above by $N_{\Phi^\pm_{BB}}^{(0)}\equiv N_{\Phi^\pm_{BB}}^{(0,0)}$. The (partial) negativity $N_{\Phi^\pm_{BB}}^{(0)}$ can be read off from (\ref{N0011})
\begin{equation}\label{N001}
N_{\Phi^\pm_{BB}}^{(0)}= \frac{1}{2}\frac{1}{\left(Z_B^\omega\right)^2}\frac{1}{\left(Z_B^\Omega\right)^2}\, \gamma_{\Phi^\pm_{BB}}\left(n_B^\omega,\, n_B^\Omega\right),
\end{equation}  
where $\gamma_{\Phi^\pm_{BB}}$ is some kind of ``cut-off function''. As all $N_{\Phi^\pm_{BB}}^{(n, m)}$ are bounded from above by $N_{\Phi^\pm_{BB}}^{(0)}$, it follows that $N_{\Phi^\pm_{BB}}$ vanishes for the same parameters as $N_{\Phi^\pm_{BB}}^{(0)}$ does. These parameters are characterized by $n_B^\omega  n_B^\Omega=1$; i.e., as soon as this fraction of the population is excited to the first state above the vacuum, state $\Phi^\pm_{BB}$ loses its entanglement. 

This can be understood in an intuitive picture considering an effective state represented by an effective density matrix $\rho^{\Phi^\pm_{BB}}_{eff}(k)$ of the $k$th sector.  As we saw, the block diagonal nature of the partially transposed reduced density matrix $ \rho_{\Phi^\pm_{BB}}^{pT}$  leads to a negativity of the form $N_{\Phi^\pm_{BB}}=N_{\Phi^\pm_{BB}}^{(0)}+\sum_{n=1}^\infty N_{\Phi^\pm_{BB}}^{(n,0)}+\sum_{m=1}^\infty N_{\Phi^\pm_{BB}}^{(0,m)} $. So we introduce $\rho^{\Phi^\pm_{BB}}_{eff} (k)$ such that the negative eigenvalue of $\left(\rho^{\Phi^\pm_{BB}}_{eff}\right)^{pT} (k)$ provides $N_{\Phi^\pm_{BB}}^{(k,0)}$. Although there is not a strict symmetry between $n$ and $m$ in (\ref{N0011}), the  effective description  captures the essential features of the behavior of entanglement, as the vanishing of $N_{\Phi^\pm_{BB}}^{(k,0)}$ implies that $N_{\Phi^\pm_{BB}}^{(0,k)}$ is vanishing as well. Imagine that only one observer is accelerated ($a_\omega\neq 0$), while the other one is inertial ($a_\Omega=0$); then we can write an effective state $\rho^\omega_{eff}(k)$ as
\begin{align}\label{effstate1}
\rho^\omega_{eff}(k)=&\alpha_k(a_\omega)| 0 k\rangle\langle 1 (k+1)|+\nonumber\\
+&\beta_k(a_\omega) |  0  (k+1)\rangle\langle 0 (k+1)|+\nonumber\\
+& \delta_k(a_\omega) | 1 k\rangle\langle 1 k|+\nonumber\\
+&\gamma_k(a_\omega) | 1 (k+1)\rangle\langle 0 k|,
\end{align}
where $\delta_k(a_\omega)\equiv 0$ and we denoted $|n_\Omega  \rangle_I\otimes|m_\omega  \rangle_I$ by $| n m\rangle$. The coherences that are present in the initial state and are responsible for entanglement are quantified by $\alpha_0(a_\omega)$ and $\gamma_0(a_\omega)$. The coefficients $\beta_k(a_\omega)$ and $\delta_k(a_\omega)$ have the physical interpretation of quantifying the occupation of the states $ |  0  (k+1)\rangle$ and $| 1 k\rangle$, respectively. Note that the Unruh effect drives the occupation of these states [in the present case only the occupation of $|  0  (k+1)\rangle$]. Initially, $\beta_k(a_\omega)=\delta_k(a_\omega)=0$. Now it is easy to see that, for fixed $k$, (\ref{effstate1}) is always entangled for finite acceleration, but  loses its entanglement for $a_\omega\to \infty$, as in this limit $\alpha_k, \beta_k, \gamma_k\to 0$ and thus $\alpha_k=\beta_k=\delta_k=\gamma_k= 0$. This explains why, for one accelerated observer (like in \cite{fuentes2005alice}), entanglement vanishes in the infinite acceleration limit but not for finite accelerations.

Moving to the general case of $a_\omega\neq 0$, $a_\Omega\neq 0$, the effective density matrix of the $k$-excitation sector is of the form
\begin{align}\label{effstate2}
\rho^{\Phi^\pm_{BB}}_{eff}(k)=&\alpha_k(a_\omega, a_\Omega)| 0 k\rangle\langle 1 (k+1)|+\nonumber\\
+&\beta_k(a_\omega, a_\Omega) |  0  (k+1)\rangle\langle 0 (k+1)|+\nonumber\\
+& \delta_k(a_\omega, a_\Omega) | 1 k\rangle\langle 1 k|+\nonumber\\
+&\gamma_k(a_\omega, a_\Omega) | 1 (k+1)\rangle\langle 0 k|
\end{align}
and the negativity vanishes for finite acceleration; i.e., $\alpha_k=\beta_k=\delta_k=\gamma_k\neq 0$ for $a_\omega, a_\Omega<\infty$. The equality between the strength of the coherences and the occupation of states $ |  0  (k+1)\rangle$ and $| 1 k\rangle$, i.e.\ $\alpha_k=\beta_k=\delta_k=\gamma_k$, is achieved due to the special structure of the reduced density matrix $\rho_{\Phi^\pm_{BB}}$ [cf. (\ref{densitymatbb})]. 

Let us have a look at the $k=0$ sector, where entanglement is initialized. For vanishing acceleration there are the coherences $ |0 0 \rangle\langle 1 1|$ and $ |1 1 \rangle \langle 0 0|$ that are non-vanishing, while the states $ | 1 0\rangle$ and $ | 0 1\rangle$ are not occupied, i.e., $\beta_0= \delta_0=0$; this state is maximally entangled. By increasing the acceleration  the symmetry of the term $a_m^2 a_n^2 |n m \rangle \langle n m|$ in the reduced density matrix $\rho_{\Phi^\pm_{BB}}$ [cf. (\ref{densitymatbb})] leads to an equal occupation of  $ | 1 0\rangle\langle 1 0|$ and $|  0  1\rangle\langle 0 1|$. At the same time the coherences  $ |0 0 \rangle\langle 1 1|$ and $ |1 1 \rangle \langle 0 0|$ are decreasing symmetrically. So for some finite acceleration $\alpha_0=\beta_0=\delta_0=\gamma_0\neq 0$ holds and entanglement vanishes.

However, there is also entanglement creation in sectors of higher excitations $k>0$ that are initially unoccupied in the sense of $\alpha_k=\beta_k=\delta_k=\gamma_k= 0$. Although entanglement is initially increasing in these sectors due to acceleration, it is vanishing for the same  acceleration as in the $k=0$ sector. This is due to the fact that, besides $a_m^2 a_n^2 |n m \rangle \langle n m|$, also $\bar{a}_m^2 \bar{a}_n^2 |(n+1) (m+1) \rangle \langle (n+1) (m+1)| $ [cf. (\ref{densitymatbb})] drives the occupation of $| 1 k\rangle\langle 1 k|$  and therefore compensates part of the loss of occupation of that state that is caused by the acceleration. This might be called ``diagonal mixing'' and is essential for achieving $\alpha_k\gamma_k=\beta_k\delta_k$ for finite acceleration. This condition is satisfied when $n_B^\omega  n_B^\Omega=1$ holds. The fact that both modes ``smear out'' due to the acceleration enables $\delta_k\neq 0$. This is the crucial point that enables the complete loss of entanglement for a finite acceleration.

 Before moving on, we want to emphasize the dependence of the negativity on the energy of the modes. The condition for entanglement is given by
\begin{equation}\label{conditionbb}
e^{-\frac{\omega }{T_\omega}}+e^{-\frac{\Omega }{T_\Omega}}\leq 1
\end{equation}
and we see that entanglement is more persistent for higher frequency modes. We introduced the Unruh temperature $T_{\omega/\Omega}=\frac{a_{\omega/\Omega}}{2\pi}$.
It is interesting to note that we can write condition (\ref{conditionbb}), for the same acceleration for both modes, i.e., $a_\omega=a_\Omega=a$, equivalently as
\begin{equation}\label{helmholtzbb}
\omega\geq T \log\left(Z_B^\Omega\right)=-F_\Omega,
\end{equation}
where $T_{}=\frac{a_{}}{2\pi}$ and $F_\Omega=-T \log\left(Z_B^\Omega\right)$ is the Helmholtz free energy. Note that the same condition with $\omega$ and $\Omega$ interchanged also holds. So at least formally the Helmholtz free energy of one mode bounds the energy (frequency) of the other one.

To summarize, by increasing the acceleration, i.e., by scanning through the families of states, entanglement decreases for all bosonic Bell states. However, there is also entanglement created in sectors that have not been entangled initially. State $\Phi^\pm_{BB}$  loses all its entanglement for a finite value of the acceleration, whereas $\Psi^\pm_{BB}$ is entangled for all finite accelerations. This is due to the appearance of the function $\gamma_{\Phi^\pm_{BB}}$ that indicates  the presence of a threshold, where the state becomes nonentangled.  The reason for this behavior is the different occupation patterns of the constituents (structures) of the states we considered here. The negativities for the states are plotted in Fig.\ \ref{picbb}. In contrast to the fermion case (cf. Sec.\ \ref{ffe}), all states lose their entanglement in the infinite acceleration limit. This is due to the infinite tower of excitations for bosonic modes that leads to a partition function $Z_B$ that grows unbounded.
Intuitively speaking, the acceleration leads to a temperature that shifts the occupation to higher energy states and therefore the occupation of the lowest lying states approaches zero. Our findings provide evidence that the structure of the states plays an important role, as this decides about the set of states that are accessible. The ``noise'' introduced by the Unruh effect is state dependent.

Next, after having addressed the bosonic case, we investigate the degradation of entanglement between a bosonic and a fermionic mode due to acceleration.

\section{Entanglement of uniformly accelerated boson-fermion states}\label{bfe}

Using the same techniques as in Secs.\ \ref{ffe} and \ref{bbe}, we study the degradation of entanglement in boson-fermion states. We start by considering the non-Bell states 
\begin{subequations}
\begin{align}
|X_1\rangle=&\frac{1}{\sqrt{2}}\left(|0_\omega\rangle_U|1^F_\Omega\rangle_U^+ +|1_\omega\rangle_U|1^F_\Omega\rangle_U^-\right),\\
|X_2\rangle=&\frac{1}{\sqrt{2}}\left(|1_\omega\rangle_U^+|1^F_\Omega\rangle_U^-+|1_\omega\rangle_U^-|1^F_\Omega\rangle_U^+\right),
\end{align}
\end{subequations}  
where $F$ labels the fermionic mode, $\omega$, $\Omega$ are the frequencies, and $0$, $1$ the occupation numbers of the Unruh modes. $+$ and $-$ refer to particles and antiparticles, respectively. The mode of frequency $\omega$ is bosonic while the mode of frequency $\Omega$ is fermionic. The respective acceleration parameters are given by $r=\text{arctanh} (e^{-\frac{\pi \omega}{a_\omega}})$ for the bosonic and $r_f=\arctan (e^{-\frac{\pi \Omega}{a_\Omega}})$ for the fermionic mode. 

Again we use the negativity (\ref{negativity}) as a measure of entanglement and obtain
\begin{align}
N_{X_1}=& 2 N_{f} N_{b, 1},\\
N_{X_2}=& 2 N_{f} N_{b, 2},
\end{align}
where $N_{f}$ is the (universal) negativity that was found for maximally entangled  fermions $N_{f}=\frac{1}{2}\cos ^2(r_f) =\frac{1}{2}(Z_F^{\Omega})^{-1}$ \cite{martin2009fermionic} and $N_{b, 1}$, $N_{b, 2}$ are given by
\begin{align}
N_{b, 1}=&\frac{1}{2}\frac{1}{\left(Z_B^\omega\right)^2}+\sum_{n=1}^\infty N_n,\\
N_{b, 2}=&\frac{1}{2}\frac{1}{Z_B^\omega}.
\end{align}
These are the negativities in the case that only  the bosons are accelerated (Appendix \ref{appbb}). Details of the calculations, as well as the expression for $N_n$, can be found in Appendix \ref{appbf}. 

Thus, the degradation of entanglement in states $X_1$ and $X_2$ is quite similar to the behavior reported in \cite{fuentes2005alice}. Intuitively, what happens is the following. When accelerated the fermions get ``rotated'' and the bosons ``smeared out''. Therefore, the fermions that are less affected by acceleration ``mimic'' the nonaccelerated bosons.  On the level of the partially transposed reduced density matrices, we observe that the fermionic and the bosonic part factorize and thus the resulting negativity can be expressed in terms of negativities obtained from the cases of one accelerated observer. 

However, as we will see, this is not a generic feature and it is absent in cases of the boson-fermion Bell states $\Psi^\pm_{BF}$ and $\Phi^\pm_{BF}$ that are given by
\begin{subequations}\label{bellfb}
\begin{align}
|\Psi^\pm_{BF}\rangle=&\frac{1}{\sqrt{2}}\left(|1_\omega\rangle_U|0^F_\Omega\rangle_U\pm|0_\omega\rangle_U|1^F_\Omega\rangle_U^+\right),\\
|\Phi^\pm_{BF}\rangle=&\frac{1}{\sqrt{2}}\left(|0_\omega\rangle_U|0^F_\Omega\rangle_U\pm|1_\omega\rangle_U|1^F_\Omega\rangle_U^+\right).
\end{align}
\end{subequations} 
The negativities of states (\ref{bellfb}) are of the form $N=\sum_{n} N_{n}$ and again each of the $N_{n}$ is bounded from above by $N_{0}$. Remember that $N_{0}$ measures the negativity in the sector, where the entanglement is initialized. In the following we denote $N_{0}$ by $N_{\Psi^\pm_{BF}/\Phi^\pm_{BF}}^{(0)}$. The further $N_{n}$ for $n\neq 0$ can be obtained analytically (Appendix \ref{appbf}). However, already with the  expression for $N_{\Psi^\pm_{BF}/\Phi^\pm_{BF}}^{(0)}$ in hand, we are able to characterize $N_{\Psi^\pm_{BF}/\Phi^\pm_{BF}}$. For states $\Psi^\pm_{BF}$ and $\Phi^\pm_{BF}$ we obtain
\begin{align}
N_{\Psi^\pm_{BF}}=&N_{\Psi^\pm_{BF}}^{(0)}+\sum_{n=1}^\infty N_{\Psi^\pm_{BF}}^{(n)},\\
N_{\Phi^\pm_{BF}}=& N_{\Phi^\pm_{BF}}^{(0)}+\sum_{n=1}^\infty N_{\Phi^\pm_{BF}}^{(n)},
\end{align}
where
\begin{align}
N_{\Psi^\pm_{BF}}^{(0)}=&\frac{1}{2}\frac{1}{Z_F^\Omega}\frac{1}{\left(Z_B^\omega\right)^2},\label{N001bf1}\\
N_{\Phi^\pm_{BF}}^{(0)}=& \frac{1}{2}\frac{1}{Z_F^{\Omega}}\frac{1}{\left(Z_B^\omega\right)^2} \, \gamma_{\Phi^\pm_{BF}}(n_B^\omega,n_F^\Omega).\label{N001bf}
\end{align}
Further, $Z_B^\omega$, $Z_F^\Omega$ are the partition functions  (\ref{partbos}), (\ref{partfer}), $n_B^\omega=(e^{\frac{\omega }{T_\omega}}-1)^{-1}$ is the Bose-Einstein distribution,  $n_F^\Omega=(e^{\frac{\Omega }{T_\Omega}}+1)^{-1}$ is the Fermi-Dirac distribution, and the $T_{\omega/\Omega}$ are the Unruh temperatures introduced by the acceleration. The function $\gamma_{\Phi^\pm_{BF}}$ is given by
\begin{equation}
\gamma_{\Phi^\pm_{BF}}=\sqrt{\frac{n_B^\omega}{n_F^\Omega}}-n_B^\omega.
\end{equation}
Details of the calculations and the expressions for $N_{\Phi^\pm_{BF}}^{(n)}$ and $N_{\Psi^\pm_{BF}}^{(n)}$ can be found in Appendix \ref{appbf}. The $N_{\Psi^\pm_{BF}/\Phi^\pm_{BF}}^{(0)}$ bound all the $N_{\Psi^\pm_{BF}/\Phi^\pm_{BF}}^{(n)}$  from above and therefore capture the essential behavior of entanglement degradation. So in the following we restrict our discussion to these quantities and refer to them as negativity.

In case of state $\Psi^\pm_{BF}$, the negativity is given by a product of the inverse partition functions for fermions and bosons. So, the negativity vanishes in the infinite acceleration limit due to the unboundedness of the bosonic partition function. As for $\Psi^\pm_{BB}$ in Sec.\ \ref{bbe}, the reason why $N_{\Psi^\pm_{BF}}^{(0)}$ is positive definite for finite accelerations is that there are no contributions to the density matrix of the form $|0 0\rangle\langle 0 0|$, since these cannot be created by the Unruh effect for states $\Psi^\pm_{}$.
\begin{figure}
\center
\includegraphics[width=\columnwidth]{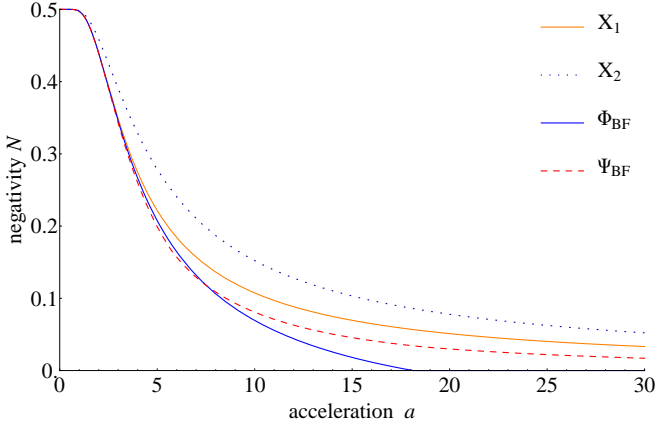}
\caption{(Color online) Negativities for the maximally entangled boson-fermion states $X_1^{}$, $X_2^{}$, $\Phi^\pm_{BF}$, and $\Psi^\pm_{BF}$, where both observers are accelerated,  plotted  against the acceleration  $a=a_\omega=a_\Omega$, measured in units of  $\frac{1}{L}$  (for some length scale $L$), for frequencies $\omega=\Omega=\frac{1}{L}$. The degradation of entanglement occurs at different rates. The negativity of state $\Phi^\pm_{BF}$ (blue continuous line) vanishes for finite accelerations, while the negativities of $X_1^{}$ (red dashed line), $X_2^{}$ (purple, dotted), and $\Psi^\pm_{BF}$ (green continuous line) vanish asymptotically.}
\label{picbf}
\end{figure}

Moving to state $\Phi^\pm_{BF}$,  we realize that, similarly to (\ref{N001}), the negativity of state $\Phi^\pm_{BF}$ vanishes for finite accelerations. Furthermore, as in (\ref{N001}), the threshold  depends on the occupation numbers of the excited modes $n_B^\omega$ and $n_F^\Omega$. When the product $n_B^\omega n_F^\Omega$ equals 1, the negativity of state $\Phi^\pm_{BF}$ vanishes. So the threshold condition is of the same form as for $\Phi^\pm_{BB}$, where it is given by $n_B^\omega n_B^\Omega=1$. However, why does the negativity of $\Phi^\pm_{BF}$ vanish while states $X_1$, $X_2$, and $\Psi^\pm_{BF}$ are entangled for all finite accelerations? First, we note that if either $r$ or $r_f$ is vanishing, $\Phi^\pm_{BF}$ is entangled for all finite accelerations. To find the answer for the generic case, we use
 an effective state description, as in Sec.\ \ref{bbe}, by
\begin{equation}\label{effstate3total}
\rho^{\Phi^\pm_{BF}}_{eff}(k)=\rho^{\Phi^\pm_{BF}}_{eff,1}(k)+e^{-\frac{\Omega}{T_\Omega}} \,\rho^{\Phi^\pm_{BF}}_{eff,2}(k),
\end{equation}
where
\begin{align}\label{effstate3}
\rho^{\Phi^\pm_{BF}}_{eff,1}(k)=&\alpha_k(a_\omega, a_\Omega)| k 0^F\rangle\langle (k+1) 1^{F+}|+\nonumber\\
+&\beta_k(a_\omega, a_\Omega) |  (k+1)  0^{F}\rangle\langle (k+1)  0^{F}|+\nonumber\\
+& \delta_k(a_\omega, a_\Omega) |  k  1^{F+}\rangle\langle k 1^{F+}|+\nonumber\\
+&\gamma_k(a_\omega, a_\Omega) | (k+1) 1^{F+}\rangle\langle k 0^F|
\end{align}
and
\begin{align}\label{effstate31}
\rho^{\Phi^\pm_{BF}}_{eff,2}(k)=&\alpha_k(a_\omega, a_\Omega)| k 1^{F-}\rangle\langle (k+1) 1^{F+}1^{F-}|+\nonumber\\
+&\beta_k(a_\omega, a_\Omega) |  (k+1)  1^{F-}\rangle\langle (k+1)  1^{F-}|+\nonumber\\
+& \delta_k(a_\omega, a_\Omega) |  k  1^{F+}1^{F-}\rangle\langle k 1^{F+}1^{F-}|+\nonumber\\
+&\gamma_k(a_\omega, a_\Omega) | (k+1) 1^{F+}1^{F-}\rangle\langle k 1^{F-}|,
\end{align}
where we denoted $|n_\omega\rangle_I\otimes| 0_\Omega \rangle_I^+  \otimes | 0_\Omega\rangle_{I}^-$ by $|  n  0^F\rangle$, $|n_\omega\rangle_I\otimes| 1_\Omega \rangle_I^+  \otimes | 0_\Omega\rangle_{I}^-$ by $|  n  1^{F+}\rangle$, $|n_\omega\rangle_I\otimes| 0_\Omega \rangle_I^+  \otimes | 1_\Omega\rangle_{I}^-$ by $|  n  1^{F-}\rangle$, and $|n_\omega\rangle_I\otimes| 1_\Omega \rangle_I^+  \otimes | 1_\Omega\rangle_{I}^-$ by $|  n  1^{F+} 1^{F-}\rangle$. Thus, if the fermionic mode is not accelerated, $\delta_k= 0$ and $\Phi^\pm_{BF}$ is entangled for all finite accelerations. We see that the effective density matrix (\ref{effstate3total}) splits into two contributions. Further, it is easy to see that (\ref{effstate3}) and (\ref{effstate31}) carry the same entanglement. Therefore, in the following, we restrict ourselves to $\rho^{\Phi^\pm_{BF}}_{eff,1}(k)$.

Similarly to the bosonic $\Phi^\pm$ state, $\alpha_0=\beta_0=\gamma_0=\delta_0\neq 0$   for $a_\omega, a_\Omega<\infty$ is achieved due to the special structure of the reduced density matrix $\rho_{\Phi^\pm_{BF}}$ [cf. (\ref{densitymatbf})]. For the further sectors, $k>0$, $\alpha_k\gamma_k=\beta_k\delta_k\neq 0$ is enabled. For vanishing acceleration there are the initially nonvanishing coherences $ | 0 0^F\rangle\langle 1 1^{F+}|$ and $| 1 1^{F+}\rangle\langle 0 0^F| $  that decrease with increasing acceleration. By increasing the acceleration some coherences are created ($ | k 0^F\rangle\langle (k+1) 1^{F+}|$ and $| (k+1) 1^{F+}\rangle\langle k 0^F| $) and, further, the term $\frac{1}{2}\cos^4 (r_f) a_n^2  (| n 0^F\rangle \langle  n 0^F|+\tan^2 (r_f)| n 1^{F+} \rangle \langle n 1^{F+} |)$ in (\ref{densitymatbf}) leads to an increasing occupation of  $ |  k  1^{F+}\rangle\langle k 1^{F+}|$ and $ |  (k+1)  0^{F}\rangle\langle (k+1)  0^{F}|$. In contrast to the bosonic $\Phi^\pm$ state, this does not happen symmetrically. However, at some point, when $r$ and $r_f$  fulfill $n_B^\omega n_F^\Omega=1$, there is an  occupation of these two states such that $\alpha_k\gamma_k=\beta_k\delta_k\neq 0$ and entanglement vanishes.

Furthermore, as for the bosonic $\Phi^\pm$-state, the term $\frac{1}{2}\cos^2\bar{a}_n^2 |(n+1) 1^{F+}\rangle \langle (n+1) 1^{F+}|$, as well as $\frac{1}{2}\sin^2 (r_f)\cos^2 (r_f) a_n^2  | n 1^{F+} \rangle \langle n 1^{F+} |$ contribute to the occupation of $ |  (k+1)  0^{F}\rangle\langle (k+1)  0^{F}|$ and therefore compensate part of the loss of occupation of that state that is caused by  acceleration. Above we called this ``diagonal mixing''. This mixing  enables $\alpha_k\gamma_k=\beta_k\delta_k\neq 0$ and therefore entanglement vanishes for finite accelerations.

Again, we want to emphasize the dependence on the energy of the entangled modes. The condition for entanglement can be written as
\begin{equation}\label{conditionfb}
n_B^\omega n_F^\Omega\leq 1.
\end{equation}
This is equivalent to $\gamma_{\Phi^\pm_{BF}}\geq 0$.  As in Sec.\ \ref{ffe}, we see that  entanglement is more persistent for higher frequency modes. 
Furthermore, we note that, for equal accelerations ($a_\omega=a_\Omega$), condition (\ref{conditionfb}) can be written in terms of the Helmholtz free energies as
\begin{equation}
\omega+\Omega\geq T \left(\log\left(Z_B^\omega\right)-\log\left(Z_F^\Omega\right)\right)=-F_\omega+F_\Omega,
\end{equation}
where $F_{\omega/\Omega}$ denote the Helmholtz free energies. Comparing that condition to (\ref{helmholtzbb}), we see a huge similarity, as (\ref{helmholtzbb}) can be written as $\omega+\Omega\geq -F_\omega-F_\Omega$. If we would take these equations for more than just a nice rewriting, we could conjecture that the origin of the nonvanishing entanglement for fermions  is given by the fact that the condition for entanglement is $\omega+\Omega\geq F_\omega+F_\Omega$, where $F_\omega+F_\Omega\leq 0$, and thus it is trivially fulfilled for all accelerations.

To summarize, the negativities of states $X_1$ and $X_2$  factorize and we observe a product structure similar to the one obtained in Sec.\ \ref{ffe}, where the total negativity is the product of the fermion and the boson contributions. That is due to the structure of the fermion mode $\Omega$. These families of states are entangled for all finite accelerations. The negativities are given by the product of inverse bosonic and fermionic partition functions and therefore vanish in the limit of infinite acceleration. In case of the $\Phi^\pm$ and $\Psi^\pm$ states ($\Phi^\pm_{BF}$ and $\Psi^\pm_{BF}$) this does not hold any more and we observe a behavior that is similar to the one we obtained for the $\Phi^\pm$ and $\Psi^\pm$ states in Sec.\ \ref{bbe}. Again, state $\Phi^\pm_{BF}$ loses all its entanglement for finite accelerations, while state $\Psi^\pm_{BF}$ is entangled for all finite accelerations; see Fig.\ \ref{picbf}. The different behavior is  due to the different structures of the states, since only in the case of state $\Phi^\pm_{BF}$ diagonal mixing is enabled. 

So, we have seen that, for fermion-fermion, boson-boson, and boson-fermion Bell states, the degradation of entanglement does not depend on, for example,  whether the state is a singlet or a triplet, but on the structure of the particular state. It is the structure of the state that determines the fading of its entanglement.

In the following section, we summarize our findings and discuss the role of particle statistics in the degradation of entanglement.

\section{Entanglement degradation and the role of particle statistics} \label{secstat}

In this section we discuss the mechanisms behind entanglement degradation and the role of particle statistics therein. Above we discussed the fermion-fermion Bell states (\ref{statesfermi}), the boson-boson Bell states (\ref{statesbb}) and the boson-fermion Bell states (\ref{bellfb}).  Using the expressions for the negativities (\ref{negastate1}), (\ref{negastate2}),  (\ref{N002}), and (\ref{N0011}), as well as (\ref{N001bf1}) and (\ref{N001bf}), we can write the negativities of all Bell states in a compact form
 \begin{equation}
 N_{S^\pm_{XY}}^{(0)}=\frac{1}{2}\frac{1}{\left(Z_X^\omega\right)^x}\frac{1}{\left(Z_Y^\Omega\right)^y}\, \gamma_{S^\pm_{XY}},\label{negativitygeneralS}
 \end{equation}
where $S^\pm_{XY}=\Psi^\pm_{XY}, \Phi^\pm_{XY}$ denotes the entangled state, $X$, $Y$ encode the statistics of the fields (fermionic, bosonic), and $x, y$ are equal to 1 for fermions ($X, Y=F$) and equal to 2 for bosons ($X, Y=B$). The functions $\gamma_{S^\pm_{XY}}$  are given by
\begin{subequations}\label{gammapsi}
\begin{align}
\gamma_{\Psi^\pm_{BB}}=&\sqrt{Z_B^\omega Z_B^\Omega+\bar{n}_{B}^2}-\bar{n}_{B},\\
\gamma_{\Psi^\pm_{BF}}=&1,\\
\gamma_{\Psi^\pm_{FF}}=&\sqrt{Z_F^\omega Z_F^\Omega+\left(Z_F^\omega Z_F^\Omega\right)^2 \bar{n}_{F}^2}-Z_F^\omega Z_F^\Omega\bar{n}_{F}
\end{align}
\end{subequations}
for the $\Psi^\pm_{XY}$ states, where $\bar{n}_{B/F}=\frac{1}{2}(n_{B/F}^\omega+n_{B/F}^\Omega)$ is the average occupation number, and 
\begin{subequations}\label{gammaphi}
\begin{align}
\gamma_{\Phi^\pm_{BB}}=&1-n_B^\omega  n_B^\Omega,\\
\gamma_{\Phi^\pm_{BF}}=&\sqrt{\frac{n_B^\omega}{n_F^\Omega}}-n_B^\omega,\\
\gamma_{\Phi^\pm_{FF}}=&1,
\end{align}
\end{subequations}
for the $\Phi^\pm_{XY}$ states. Equation (\ref{negativitygeneralS}) gives the negativities in the $k=0$ sector, where entanglement is initiated. For some states there is entanglement dynamically created in other sectors but these are always bounded from above by (\ref{negativitygeneralS}); see Fig.\ \ref{piccreat}. Therefore, these negativities capture the main features of entanglement degradation and we restrict our attention to these.

Physically speaking, after fixing the frequencies $\omega$ and $\Omega$, as seen by the accelerated observers, Eq.\ (\ref{negativitygeneralS}) gives the negativity of the two-parameter family of states $S^\pm_{XY}$. That is, for each choice of the pair $(a_\omega, a_\Omega)$, Eq.\ (\ref{negativitygeneralS}) gives the negativity of the particular state $S^\pm_{XY}$ that is characterized by $(\omega, \Omega, a_\omega, a_\Omega)$, when this state is seen by accelerated observers of accelerations $a_\omega$ and $a_\Omega$, respectively. For this reason, and also because the Unruh modes that are considered are global modes, the setting should be considered as a toy model that captures the essential features of entanglement degradation.

     \begin{figure}
        \centering
        \begin{subfigure}[b]{0.475\columnwidth}
            \centering
            \includegraphics[width=\columnwidth]{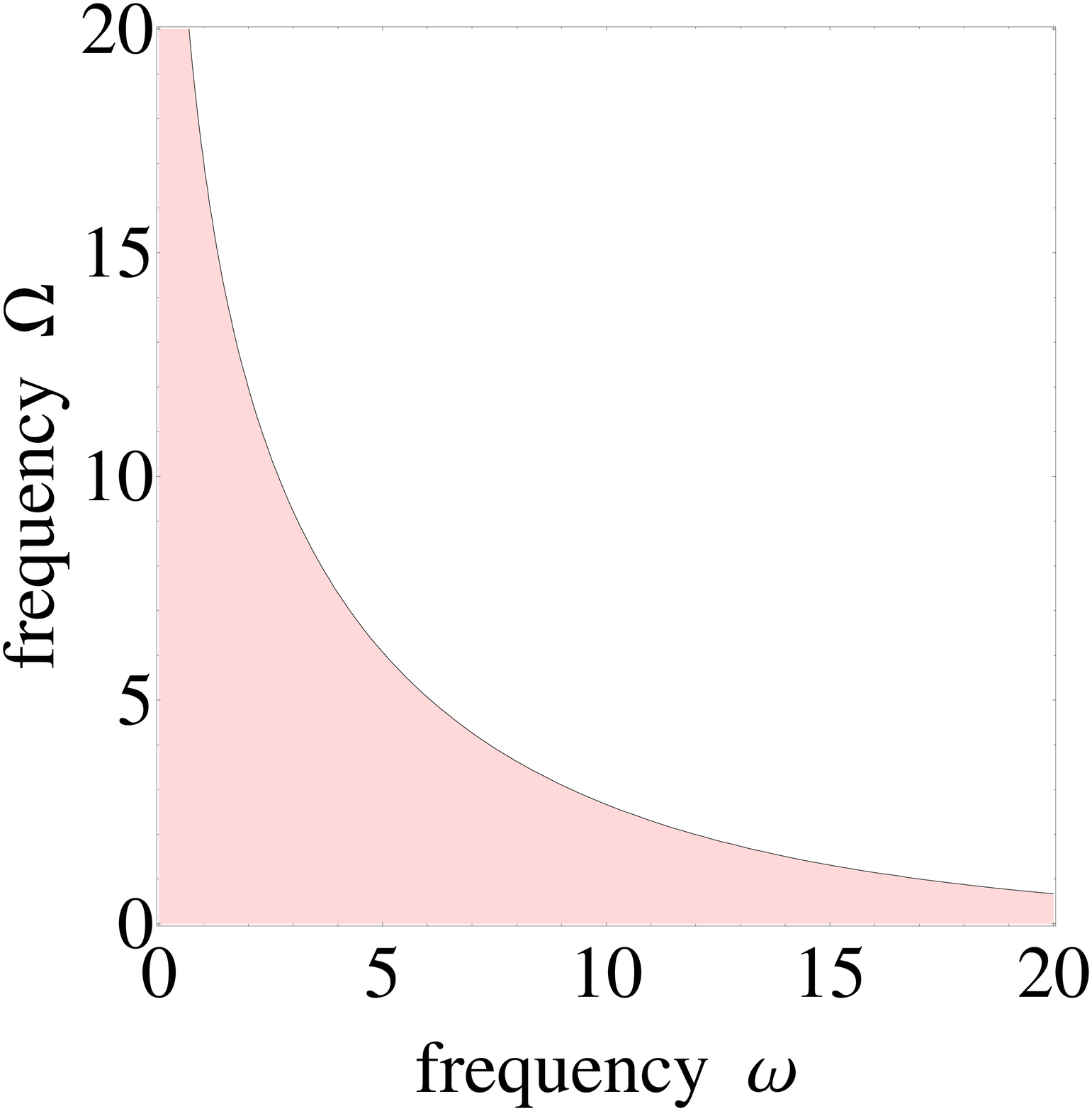}
            \caption[Network2]%
            {State $\Phi^\pm_{BB}$ for acceleration $a_\omega=a_\Omega=50\frac{1}{L}$}    
        \end{subfigure}
        \hspace{1mm}
        \begin{subfigure}[b]{0.475\columnwidth}  
            \centering 
            \includegraphics[width=\columnwidth]{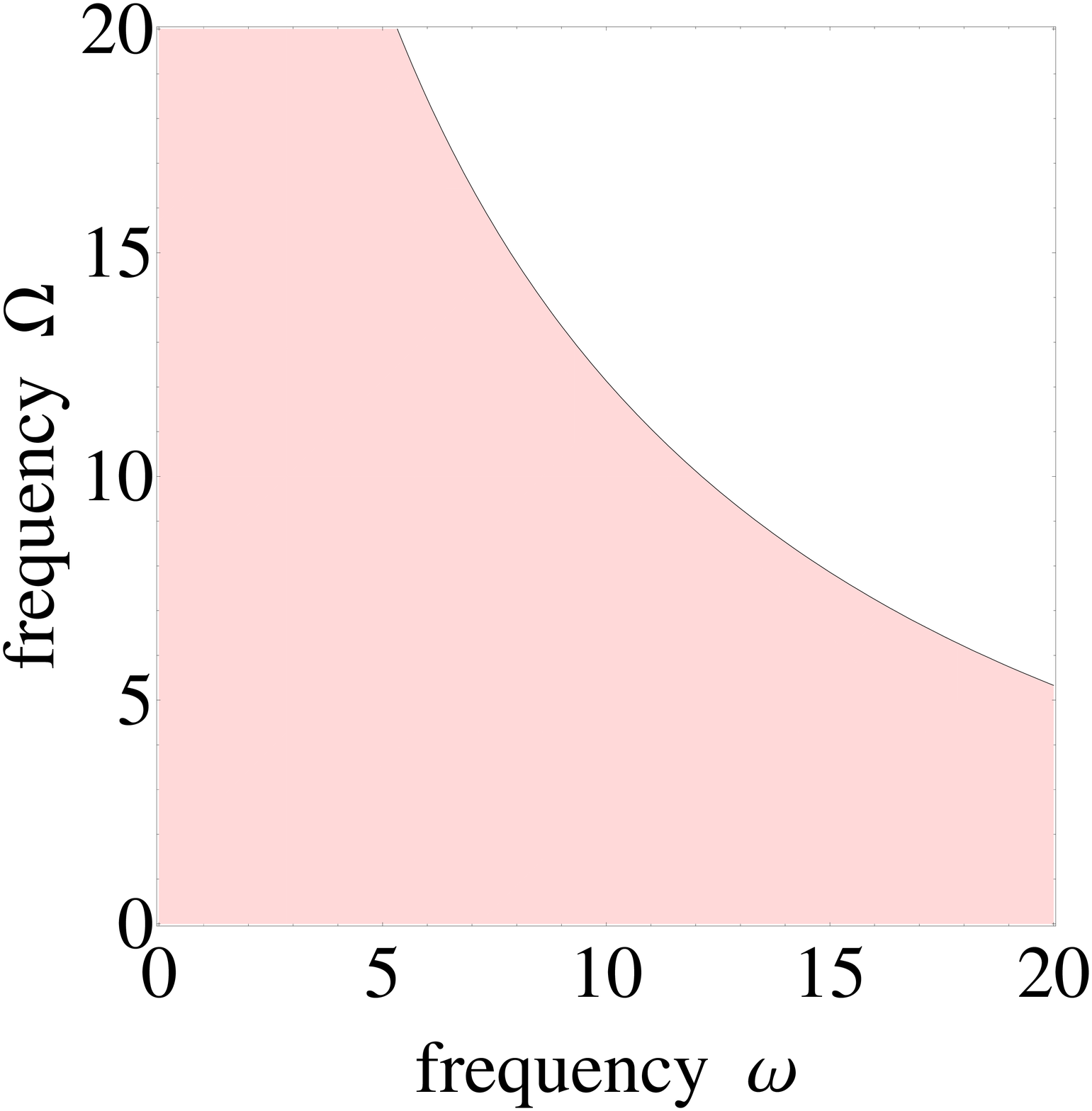}
            \caption[]%
            {State $\Phi^\pm_{BB}$ for acceleration $a_\omega=a_\Omega=100\frac{1}{L}$}    
        \end{subfigure}
        \vskip\baselineskip
        \begin{subfigure}[b]{0.475\columnwidth}   
            \centering 
            \includegraphics[width=\columnwidth]{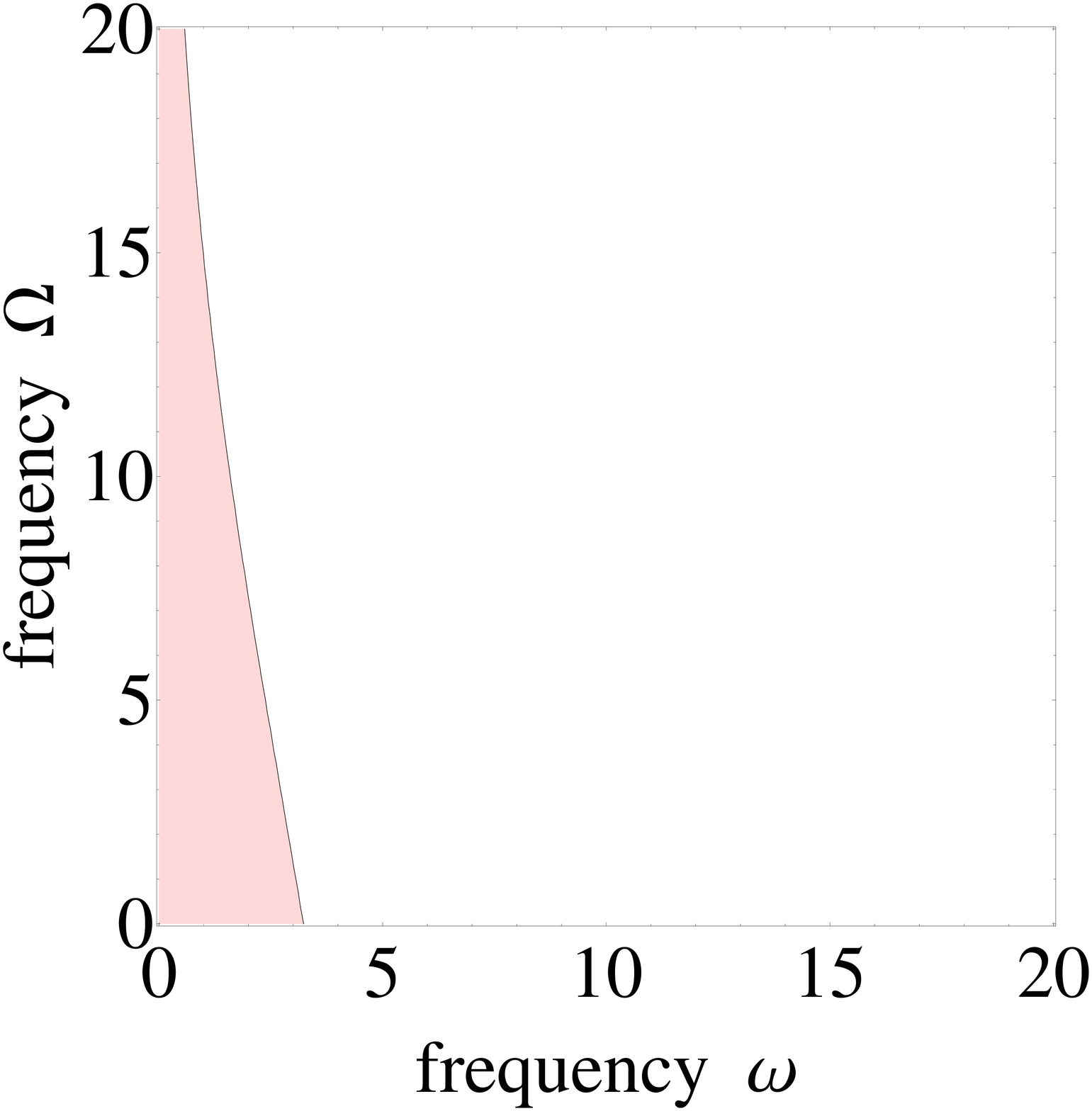}
            \caption[]%
            {State $\Phi^\pm_{BF}$ for acceleration $a_\omega=a_\Omega=50\frac{1}{L}$}    
        \end{subfigure}
        \quad
        \begin{subfigure}[b]{0.475\columnwidth}   
            \centering 
            \includegraphics[width=\columnwidth]{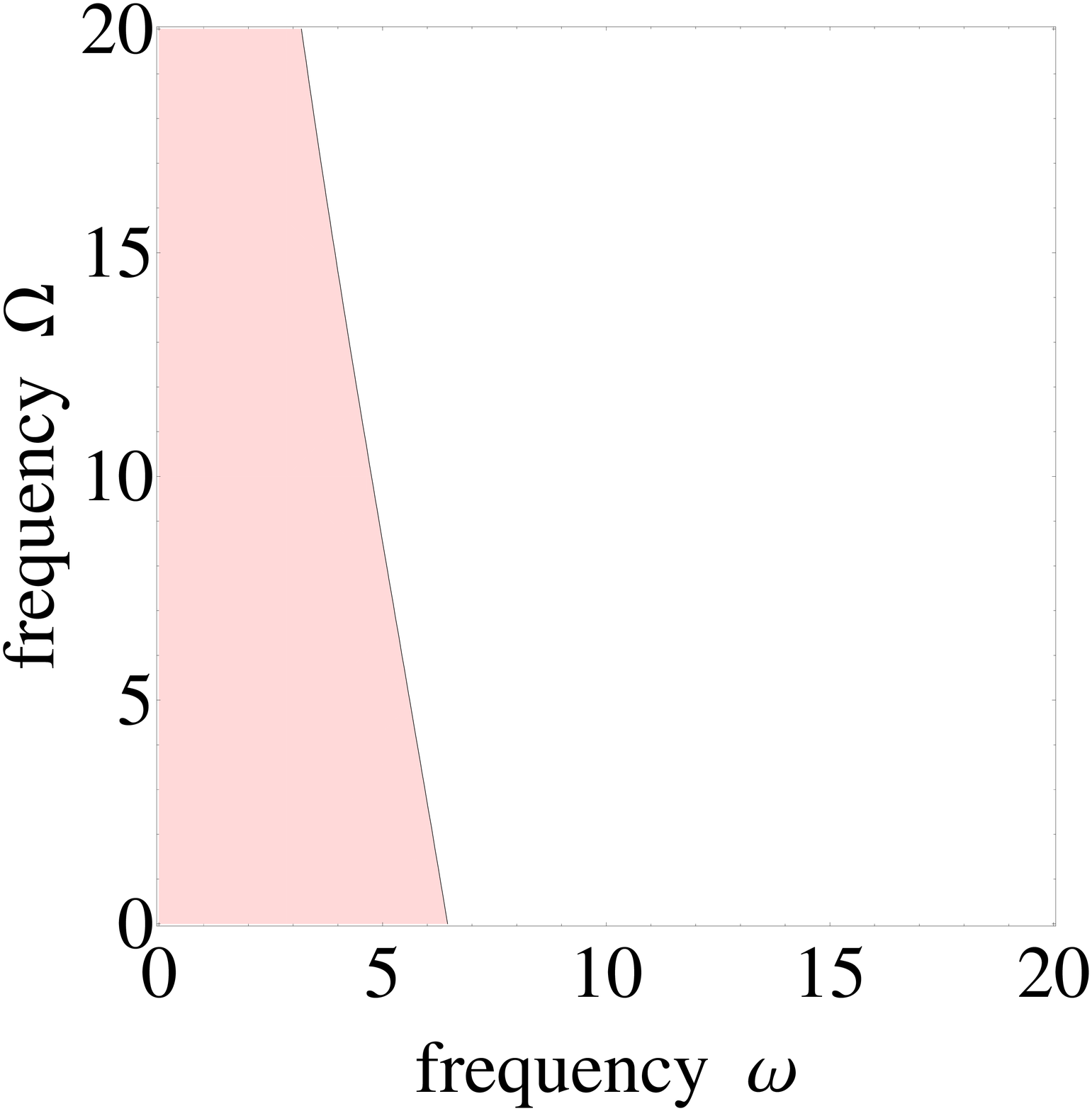}
            \caption[]%
            {State $\Phi^\pm_{BF}$ for acceleration $a_\omega=a_\Omega=100\frac{1}{L}$}    
        \end{subfigure}
        \caption
        {(Color online) Energy dependence of the entanglement in Bell states $\Phi^\pm_{BB/BF}$: States of modes of frequencies $\omega$ and $\Omega$ (in units of $\frac{1}{L}$ for some length scale $L$) contained in the red region show no entanglement, while states of higher energies remain entangled (white region). The plots show states $\Phi^\pm_{BB/BF}$ for acceleration $a_\omega=a_\Omega=50\frac{1}{L}$ ((a), (c)) and $a_\omega=a_\Omega=100\frac{1}{L}$ ((b), (d)). State $\Phi^\pm_{FF}$ is entangled for all frequencies and accelerations and therefore it is not shown. The zeros of the functions $\gamma_{\Phi^\pm_{BB}}$ and $\gamma_{\Phi^\pm_{BF}}$ define the border between the regions. The asymmetry in (c) and (d) is due to the fact that fermions are ``more resistant'' towards the effects of acceleration.} 
        \label{picentsytat}
    \end{figure}
    
There are essentially two factors determining the fading of entanglement. The first one is given by the set of states (as above, by states we mean diagonal elements of the density matrix) that become available when a state is accelerated. This set depends heavily on the structure of the state, as, for example, for $\Psi^\pm$ the state $|00\rangle\langle 0 0|$ never becomes accessible, but also on the statistics that does not allow for two- or more-particle states for fermions. The second determining factor is whether the population of states $|00\rangle\langle 0 0|$, $|01\rangle\langle 0 1|$, $|10\rangle\langle 1 0|$, and $|1 1\rangle\langle 1 1|$ can be transferred completely to higher excited states like, for example, $|2 1\rangle\langle 2 1|$. If that is possible also the coherences $|00\rangle\langle 1 1|$ and $|01\rangle\langle 1 0|$ vanish with increasing acceleration. Both factors depend heavily on the statistics of the underlying field. 

For illustrating reasons, consider the density matrix $\rho$,
\begin{equation}\label{ilus}
\rho=\left(
\begin{array}{cccc}
 \rho_{0000}^{\Phi}&   0& 0& \rho_{0011}^{\Phi} \\
 0 &   \rho_{0101}^{\Psi}& \rho_{0110}^{\Psi}& 0 \\
 0& \rho_{1001}^{\Psi}& \rho_{1010}^{\Psi}& 0\\
 \rho_{1100}^{\Phi}& 0& 0& \rho_{1111}^{\Phi}\\
\end{array}
\right),
\end{equation}
written in the basis $\{|00\rangle, |01\rangle, |10\rangle, |11\rangle\}$, where all $\rho_{ijkl}^{\Phi}$ ($i, j, k, l \in \{0, 1\}$) are zero for $\Psi^\pm$ states and vice versa. Tracing out anti-particles, (\ref{ilus}) is the full density matrix for fermions and acceleration decreases  the diagonal elements ($\rho_{0000}^{\Phi}$, $\rho_{1111}^{\Phi}$; $\rho_{0101}^{\Psi}$, $\rho_{1010}^{\Psi}$), as well as the coherences ($\rho_{0011}^{\Phi}$, $\rho_{1100}^{\Phi}$; $\rho_{1001}^{\Psi}$, $\rho_{0110}^{\Psi}$) to a finite value for both $\Phi^\pm$ and $\Psi^\pm$ states. Furthermore, the  ``squared'' coherences $\rho_{0011}^{\Phi} \rho_{1100}^{\Phi}$ and $\rho_{1001}^{\Psi} \rho_{0110}^{\Psi}$ always dominate the product of the occupations of states $\rho_{0000}^{\Phi}$, $\rho_{1111}^{\Phi}$ and $\rho_{0101}^{\Psi}$, $\rho_{1010}^{\Psi}$, respectively. That is why entanglement decreases but does not vanish. 

For bosons, in contrast, (\ref{ilus}) is only the initially nonvanishing part of the infinite-dimensional density matrix. The reason why $\Phi^\pm$ states lose their entanglement (cf. Fig.\  \ref{picentsytat}), while $\Psi^\pm$ are entangled for all finite accelerations, is given by the set of available states (cf. Secs.\ \ref{bbe} and \ref{bfe}). The reason for the asymptotic vanishing of the negativity for all boson-boson and boson-fermion states is given by the second determining factor (see above).  All initially nonvanishing elements of $\rho$, i.e., the $\rho_{ijkl}^{\Phi}$ and the $\rho_{ijkl}^{\Psi}$, approach zero in the infinite acceleration limit, as matrix elements of higher excitation states are increasing. That is due to bosonic statistics. Naively, fermions get ``rotated'' and bosons ``smeared'' towards higher excited sectors due to acceleration. Furthermore, note the asymmetry in Fig.\  \ref{picentsytat} that is due to the fact that fermions are ``more resistant'' towards the effects of acceleration.

This explanation based on the two determining factors captures the role of particle statistics. Particle statistics also is reflected in the negativities (\ref{negativitygeneralS}) that are written in terms of the partition functions and the occupation numbers and therefore make the ``effects of statistics'' apparent.

In this work we studied maximally entangled states like, for example,
\begin{subequations}\label{statesbbalpha}
\begin{align}
|\Psi_{\alpha}\rangle=&\sin (\alpha)|0_\omega\rangle_U|1_\Omega\rangle_U+\cos (\alpha)|1_\omega\rangle_U|0_\Omega\rangle_U,\\
|\Phi_{\alpha}\rangle=&\sin (\alpha)|0_\omega\rangle_U|0_\Omega\rangle_U+\cos (\alpha)|1_\omega\rangle_U|1_\Omega\rangle_U,
\end{align}
\end{subequations}
where we chose $\alpha$ to be $\pi/4$.  Differing choices of $\alpha$ lead to less entangled states, as the negativity $N_{\Psi_{\alpha}}=N_{\Phi_{\alpha}}=\sin (\alpha)\cos (\alpha)$ is maximized for $\alpha=\pi/4$. Given the mechanisms that lead to entanglement degradation that we outlined above, it is evident that states (\ref{statesbbalpha}) behave qualitatively  the same for generic $\alpha$ as they do for $\alpha=\pi/4$, i.e., in the maximally entangled case. In the case of more general mixed states, we expect that, depending on the state, one observes that the degradation of entanglement shows characteristics that are best described by a mixture of the characteristics of the degradations in the case of $\Phi^\pm$  and $\Psi^\pm$. However, it seems reasonable to expect that a random mixed state will lose its entanglement for finite acceleration with high probability, as in the fermionic case studied in \cite{montero2011fermionic2}.

In some sense we can think of the functions $\gamma$ [(\ref{gammapsi}) and (\ref{gammaphi})] as deformations of a (universal) particle-statistics-dependent negativity $\mathcal{N}_U$,
\begin{equation}
\mathcal{N}_U=\frac{1}{2}\frac{1}{\left(Z_X^\omega\right)^x}\frac{1}{\left(Z_Y^\Omega\right)^y},
\end{equation}
that only depends on the partition functions that are characteristic for the particle statistics of the field. Then the particular structure of the Bell state, as well as the particle statistics, set the $\gamma$ that we might call structure functions and denote them by $\gamma_{structure}$. As we saw, these depend heavily on the set of states whose occupation is driven by the Unruh effect. So finally we can write the entanglement of a Bell state (negativity $N_{state}$) in the sector where  entanglement is initialized as
\begin{equation}\label{schema}
N_{state}=\gamma_{structure}\,\mathcal{N}_U.
\end{equation}

We want to close this section by giving some comments on Eq.\ (\ref{schema}). First, choosing the partitioning of $N_{state}$ in $\gamma_{structure}$ and $\mathcal{N}_U$ is not unique and there are  possible partitions different from (\ref{negativitygeneralS}). 	Nevertheless, writing the negativity in  form (\ref{schema}) makes the importance of the particular structure of the state manifest. Moreover, Eq.\ (\ref{schema}) makes it possible to clearly identify the two determining factors: The first one sets the function $\gamma_{structure}$, while the second one determines $\mathcal{N}_U$. Further, we note that a slightly varied form of (\ref{schema}) also holds for states $X_1$ and $X_2$. We expect that there are slight modifications when there are different particles involved, like particles carrying spin. Finally, it would be interesting to figure out whether the negativities of non-maximally entangled mixed states could also be captured in an expression similar to  (\ref{schema}).

In the following section, we briefly point out possible implications of the above findings for particles in Bell states close to the black hole horizon.

\section{Degradation of entanglement in the vicinity of a black hole} \label{secbh}


The framework we used in this work also applies to the spacetime close to a black hole, as outlined in Appendix \ref{appbh} (see also \cite{martin2010unveiling}). However, given the caveats in the interpretation of the states, we described in Sec.\ \ref{ffe}, the following discussion aims at giving a qualitative idea about entanglement degradation near black holes. 

One point that can be inferred is that entanglement gets degraded in the vicinity of the black hole horizon. Further, all states that involve bosons lose their entanglement in the limit of reaching the horizon. This is in contrast to the fermion-fermion states, where entanglement never vanishes. Further, there are crucial differences between the degradation of entanglement for states $\Phi^\pm_{}$ and $\Psi^\pm_{}$. The entanglement of states $\Phi^\pm_{BB}$ and $\Phi^\pm_{BF}$  completely vanishes at a finite distance from the horizon that is large compared to the Planck length $L_P$ ($d\approx 0.01 R_S$, where $R_S$ is the Schwarzschild radius), whereas states $\Psi^\pm_{}$ are entangled for any finite distance from the black hole.

Thus, we observed that entanglement, an important resource for quantum information tasks, gets degraded very differently for differing Bell states, i.e., the degradation of entanglement is state dependent. Our findings imply that there are particular states that remain entangled as seen by an observer that is uniformly accelerated or equivalently is stationary close to a black hole, while, for other choices of the state, there is no entanglement remaining. This implies that the gravitational degradation of entanglement depends on the structure of the state.

\section{Conclusions} \label{conclusions}

In this work, we studied families of two uniformly accelerated maximally entangled Unruh modes in the general case of different accelerations and analyzed their entanglement, measured by the negativity. Therefore, we considered states containing two fermionic modes, two bosonic modes, as well as states  of one bosonic and one fermionic mode. Special emphasis was given to the comparison of Bell states $\Phi^\pm$ and $\Psi^\pm$. Although the Unruh modes we used do not have a simple physical interpretation, our studies provide insight into the mechanisms that lead to the degradation of entanglement due to acceleration. 

We found that, in contrast to the other cases, purely fermionic families of Bell states are entangled for all accelerations. Still, the entanglement of state $\Psi^\pm$ degrades faster with acceleration than the entanglement of state $\Phi^\pm$. Interestingly, it is only for state $\Psi^\pm$  that both accelerated modes give rise to a contribution to the same diagonal element of the reduced density matrix that is relevant for entanglement. We suspect that this special feature of $\Psi^\pm$ is responsible for the different behavior regarding entanglement degradation. Furthermore, we found that also classical correlations are partially lost due to acceleration.

In the purely bosonic case, as well as in the boson-fermion case, state $\Psi^\pm$ remains entangled for all finite accelerations, and entanglement vanishes asymptotically  in the limit of infinite accelerations. In contrast,  state $\Phi^\pm$ loses its entanglement for some finite acceleration. This is manifest in the presence of a ``cut-off function'' $\gamma_{\Phi^\pm_{}}$ in the expression for the negativity. So we found that the type of Bell state (i.e., being $\Phi^\pm$ or $\Psi^\pm$) crucially affects the robustness of its entanglement against acceleration. Furthermore, we obtained that the reason for the occurrence of this phenomenon is originated in the particular occupation patterns of the constituents (the ``structure'') of the state, which determine which excitations can be driven by the Unruh effect. 

Applying an effective state picture, we were able to explain this crucial difference between both types of states. State $\Psi^\pm$ is entangled for all finite accelerations as the Unruh effect does not drive the occupation of state $|00\rangle\langle 0 0 |$, and this state is naturally absent in the density matrix of  $\Psi^\pm$. Entanglement vanishes only  asymptotically for infinite acceleration, as in this regime all occupation is shifted towards highly excited states. For state $\Phi^\pm$ things are different. Essentially what happens is the following. For vanishing acceleration, there exist the coherences $ |0 0 \rangle\langle 1 1|$ and $ |1 1 \rangle \langle 0 0|$ that are responsible for the entanglement, while the occupation of $ | 1 0\rangle\langle 1 0|$ and $|  0  1\rangle\langle 0 1|$ is vanishing. When acceleration is increasing, the coherences are decreasing, while at the same time the occupation of $ | 1 0\rangle\langle 1 0|$ and $|  0  1\rangle\langle 0 1|$ is driven (symmetrically) by the Unruh effect by creating one excitation in $|00\rangle\langle 0 0 |$ . Thus, for the value of the acceleration for which the condition for entanglement [cf. (\ref{conditionbb}), (\ref{conditionfb})] is violated, entanglement vanishes. Hence, we traced the difference in the behavior regarding entanglement degradation back to the set of accessible states and the symmetry in the distribution of probability among them. It seems that diagonal mixing, as we coined it above, is required to achieve sufficient uniformity in the occupation of the states.

 Further, we found that there are two factors that determine the fading of entanglement. The first one, given by the set of states that become accessible due to the Unruh effect, is heavily influenced by the structure of the state. Said factor determines whether a state loses its entanglement for finite accelerations. The second factor is more closely related to the particle statistics of the modes that constitute the Bell state. It is whether higher excitation states become accessible due to acceleration. That is the case for bosonic modes, and thus Bell states in which these modes are involved are nonentangled in the infinite acceleration limit, whereas purely fermionic Bell states are always entangled. Remarkably, we found that the negativities of the boson-boson, boson-fermion, and fermion-fermion Bell states can be expressed in the same form (\ref{negativitygeneralS}),
\begin{equation}
N_{S^\pm_{XY}}^{(0)}=\frac{1}{2}\frac{1}{\left(Z_X^\omega\right)^x}\frac{1}{\left(Z_Y^\Omega\right)^y}\, \gamma_{S^\pm_{XY}},
\end{equation}
where the $Z_{B/F}^{\omega/\Omega}$ are the partition functions (of a harmonic oscillator or two-level system with energy gap $\omega/\Omega$) and the $\gamma_{S^\pm_{XY}}$ are functions determined by the first factor, we introduced above. 
 

Furthermore, we discussed possible effects of hovering over a black hole on entangled states of two Unruh modes.

In summary, our studies reveal the mechanisms that cause the behavior of entanglement in accelerated frames to depend heavily on the particular occupation patterns of the constituents of the entangled state.

\begin{acknowledgements}
We would like to thank Dieter L\"ust for useful discussions and Andrzej Dragan for comments and bringing references \cite{dragan2013localized, dragan2013localized2, doukas2013entanglement} to our attention. Further, we are grateful for the support from Funda\c{c}\~{a}o para a Ci\^{e}ncia e a Tecnologia (Portugal), namely through the programs PTDC/POPH and projects UID/Multi/00491/2013, UID/EEA/50008/2013, IT/QuSim and CRUP-CPU/CQVibes, partially funded by EU FEDER, and from the EU FP7 projects LANDAUER (GA 318287) and PAPETS (GA 323901). BR acknowledges the support from the DP-PMI and FCT (Portugal) through scholarship SFRH/BD/52651/2014.
\end{acknowledgements}

\appendix

\section{Fermion-fermion states}\label{appneg}

\subsection{Calculation of negativities}

The negativity for a composite system (we denote the subsystems by $A$ and $B$) described by a density matrix $\rho=\rho_{AB}$  is given by the sum of the absolute values of the negative eigenvalues of the partially transposed density matrix $\rho_{AB}^{pT}$,
\begin{equation}\label{negativity}
N=\frac{1}{2}\sum_j \left(|\lambda_j|-\lambda_j\right),
\end{equation}
where the $\lambda_j$'s are the eigenvalues of $\rho_{AB}^{pT}$ and for a density matrix $\rho_{AB}=\sum_{klmn} p_{klmn} | k \rangle\langle l|\otimes | m \rangle\langle n|$ the partial transposed is given by $\rho_{AB}^{pT}=\sum_{klmn} p_{klmn} | k\rangle\langle l|\otimes | n \rangle\langle m|$.

The calculations of the negativities of the fermion states $\{\psi_i\}=\{\Psi_{FF}^\pm, \Phi_{FF}^\pm\}$ have to be carried out with care and  the braided tensor product has to be taken into account \cite{bradler2012comment}. As above, when we introduced the Unruh modes, we chose the ordering $| ijkl \rangle_{\tilde{\omega}}=| i_{\tilde{\omega}} \rangle_I^+ \otimes | j_{\tilde{\omega}} \rangle_{II}^- \otimes | k_{\tilde{\omega}} \rangle_I^- \otimes | l_{\tilde{\omega}}\rangle_{II}^+$. The density matrices are obtained as $\rho_{I,II}^{(i)}=|\psi_i\rangle\langle\psi_i|$. To obtain the  reduced density matrices  $\rho_i^{}=Tr_{II} \left(\rho_{I,II}^{(i)}\right)$ ($i\in \{\Psi_{FF}^\pm, \Phi_{FF}^\pm\}$), we trace out modes supported in region $II$ and take care of the operator ordering. Finally, we partially transpose  the reduced density matrices and identify the blocks of $\rho_i^{pT}$ that admit negative eigenvalues.

In the case that the observers are able to detect particles as well as anti-particles, the relevant blocks $b^{\Psi_{FF}^+}_\mu$ of $\rho_{\Psi_{FF}^+}^{pT}$ are given by
\begin{align}
b^{\Psi_{FF}^+}_1=&c^{\Psi_{FF}^+}_{\epsilon=1},\\
b^{\Psi_{FF}^+}_2=&\tan^2 (r_f^\omega) c^{\Psi_{FF}^+}_{\epsilon=-1},\\
b^{\Psi_{FF}^+}_3=&\tan^2 (r_f^\Omega) c^{\Psi_{FF}^+}_{\epsilon=1},\\
b^{\Psi_{FF}^+}_4=&\tan^2 (r_f^\Omega) \tan^2 (r_f^\omega) c^{\Psi_{FF}^+}_{\epsilon=-1},
\end{align}
where
\begin{align}
c^{\Psi_{FF}^+}_\epsilon=&\frac{1}{2}\cos ^2(r_f^\omega)\cos ^2(r_f^\Omega)\nonumber \\
 \times&\left(
\begin{array}{cc}
  0 &   \epsilon\cos (r_f^\omega)\cos (r_f^\Omega) \\
 \epsilon\cos (r_f^\omega)\cos (r_f^\Omega)   &   \sin ^2(r_f^\omega)+\sin ^2(r_f^\Omega)    \\
\end{array}
\right).
\end{align}
The negativity $N_{\Psi_{FF}^+}$ is sum of the absolute values of the negative eigenvalues of $\rho_{\Psi_{FF}^+}^{pT}$. Thus, using that $N(c^{\Psi_{FF}^+}_{\epsilon=1}) =N(c^{\Psi_{FF}^+}_{\epsilon=-1})$ and the fact that the $b^{\Psi_{FF}^-}_\mu$ can be obtained from the $b^{\Psi_{FF}^+}_\mu$ by the replacement $\epsilon\to-\epsilon$, we can write
\begin{align}
N_{\Psi^\pm_{FF}}=&(1+ \tan^2 (r_f^\omega) + \tan^2 (r_f^\Omega) + \tan^2 (r_f^\omega)  \tan^2 (r_f^\Omega) )\times \nonumber\\
\times& N(c^{\Psi_{FF}^+}_{\epsilon=1}) \nonumber\\
=&\frac{1}{4}( -\left( \sin ^2(r_f^\omega)+\sin ^2(r_f^\Omega)\right)+\nonumber\\
+& \sqrt{\left( \sin ^2(r_f^\omega)+\sin ^2(r_f^\Omega)\right)^2+4\cos ^2(r_f^\omega) \cos ^2(r_f^\Omega)}).
\end{align}
As expected, assuming that only particles can be detected by the observers, i.e., tracing out anti-particles the negativity does not change. In this case, there is only one block contributing to the negativity that is given by
\begin{align}
&\frac{1}{2}\left(
\begin{array}{cc}
  0 &   \cos (r_f^\omega)\cos (r_f^\Omega) \\
 \cos (r_f^\omega)\cos (r_f^\Omega)   &   \sin ^2(r_f^\omega)+\sin ^2(r_f^\Omega)    \\
\end{array}
\right),
\end{align}
which has the negative eigenvalue $-N_{\Psi^\pm_{FF}}$ and thus entanglement remains unchanged.

Considering state $\Phi_{FF}^+$, in the case that the observers are able to detect particles as well as anti-particles, the relevant blocks $b^{\Phi_{FF}^+}_\mu$ of $\rho_{\Phi_{FF}^+}^{pT}$ are calculated as above and are given by
\begin{align}
b^{\Phi_{FF}^+}_1=&c^{\Phi_{FF}^+}_{\epsilon=1},\\
b^{\Phi_{FF}^+}_2=&\tan^2 (r_f^\omega) c^{\Phi_{FF}^+}_{\epsilon=-1},\\
b^{\Phi_{FF}^+}_3=&\tan^2 (r_f^\Omega) c^{\Phi_{FF}^+}_{\epsilon=1},\\
b^{\Phi_{FF}^+}_4=&\tan^2 (r_f^\Omega) \tan^2 (r_f^\omega) c^{\Phi_{FF}^+}_{\epsilon=-1},
\end{align}
where
\begin{align}
c^{\Phi_{FF}^+}_\epsilon=&\frac{1}{2}\cos ^2(r_f^\omega)\cos ^2(r_f^\Omega) \times\nonumber\\
 \times&\left(
\begin{array}{cc}
 \sin^2 (r_f^\omega) \cos^2 (r_f^\Omega)  &   \epsilon\cos (r_f^\omega)\cos (r_f^\Omega) \\
 \epsilon \cos (r_f^\omega)\cos (r_f^\Omega)  &   \sin^2 (r_f^\Omega) \cos^2 (r_f^\omega)  \\
\end{array}
\right).
\end{align}
We notice that $N(c^{\Phi_{FF}^+}_{\epsilon=1}) =N(c^{\Phi_{FF}^+}_{\epsilon=-1})$ and that again $b^{\Psi_{FF}^+}_\mu\to b^{\Psi_{FF}^-}_\mu$ is induced by $\epsilon\to-\epsilon$. So we  can write the negativity as
\begin{align}
N_{\Phi^\pm_{FF}}=&(1+ \tan^2 (r_f^\omega) + \tan^2 (r_f^\Omega) + \tan^2 (r_f^\omega)  \tan^2 (r_f^\Omega) )\times\nonumber\\
\times& N(c^{\Phi_{FF}^+}_{\epsilon=1}) \nonumber\\
=& \frac{1}{2}\cos ^2(r_f^\omega)\cos ^2(r_f^\Omega).
\end{align}
Assuming that the observers can only detect particles, does not affect the negativity. Tracing over anti-particles in $\rho_{\Phi_{FF}^+}^{pT}$ leads to
\begin{align}
&\frac{1}{2}\left(
\begin{array}{cc}
 \sin^2 (r_f^\omega) \cos^2 (r_f^\Omega)  &   \cos (r_f^\omega)\cos (r_f^\Omega) \\
  \cos (r_f^\omega)\cos (r_f^\Omega)  &   \sin^2 (r_f^\Omega) \cos^2 (r_f^\omega)  \\
\end{array}
\right)
\end{align}
with the negative eigenvalue $-N_{\Phi^\pm_{FF}}$.

\subsection{Entanglement in different sectors}\label{appa2}

\begin{figure}[t]
\center
\includegraphics[width=\columnwidth]{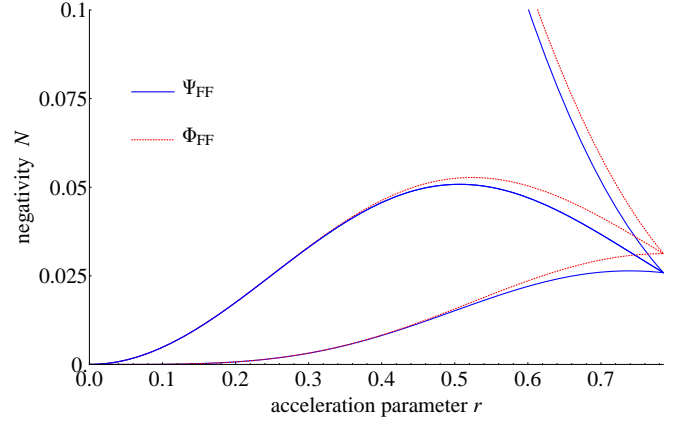}
\caption{(Color online) Entanglement distribution for states $\Phi^\pm_{FF}$ (red dotted line) and $\Psi^\pm_{FF}$ (blue continuous line) plotted versus the dimensionless acceleration parameter $r=r_f^\omega=r_f^\Omega$. The infinite acceleration limit corresponds to $r=\frac{\pi}{4}$ . The middle pair of solid curves is twofold degenerate due to the symmetry of the states. With increasing acceleration, entanglement is created in previously separable sectors. From top to bottom the curves correspond as follows: red, $N(0,0\,|\,1^+,1^+)$, $N(1^-,0\,|\,1^+1^-,1^+)$ and $N(0,1^-\,|\,1^+,1^+1^-)$, $N(1^-,1^-\,|\,1^+1^-,1^+1^-)$; blue, $N(1^+,0\,|\, 0,1^+)$, $N(1^+1^-,0\,|\, 1^-,1^+)$ and $N(1^+,1^-\,|\, 0,1^+1^-)$, $N(1^+1^-,1^-\,|\, 1^-,1^+1^-)$.}\label{vergleich}
\end{figure}

We briefly discuss the distribution of entanglement in states $\psi_i$, when the acceleration is non-vanishing. While  entanglement in the sector where it was initiated is decreasing with increasing acceleration, there is entanglement created in previously nonentangled sectors. Schematically, we can write $N_{\Phi^\pm_{FF}}$ as
\begin{align}
N_{\Phi^\pm_{FF}}=&N(0,0\,|\,1^+,1^+)\nonumber\\
+&N(1^-,0\,|\,1^+1^-,1^+)\nonumber\\
+&N(0,1^-\,|\,1^+,1^+1^-)\nonumber\\
+&N(1^-,1^-\,|\,1^+1^-,1^+1^-),
\end{align}
where
\begin{align}
&N(0,0\,|\,1^+,1^+)= N(c^{\Phi_{FF}^+}_{\epsilon=1}),\\
&N(1^-,0\,|\,1^+1^-,1^+)=\tan^2 (r_f^\omega) N(c^{\Phi_{FF}^+}_{\epsilon=1}),\label{creatneg1}\\
&N(0,1^-\,|\,1^+,1^+1^-)=\tan^2 (r_f^\Omega) N(c^{\Phi_{FF}^+}_{\epsilon=1}),\\
&N(1^-,1^-\,|\,1^+1^-,1^+1^-)=\tan^2 (r_f^\omega)\tan^2 (r_f^\Omega) N(c^{\Phi_{FF}^+}_{\epsilon=1})\label{creatneg3}
\end{align}
are the negativities of the different sectors. $N(0,0\,|\,1^+,1^+)$ is the negativity of the sector, where the entanglement is initialized, i.e., entanglement between $| 0_\omega \rangle_I^+ \otimes | 0_\omega \rangle_{I}^- \otimes | 0_\Omega \rangle_I^+ \otimes | 0_\Omega\rangle_{I}^-$ and $| 1_\omega \rangle_I^+ \otimes | 0_\omega \rangle_{I}^- \otimes | 1_\Omega \rangle_I^+ \otimes | 0_\Omega\rangle_{I}^-$. There are three sectors in which entanglement is created due to acceleration (\ref{creatneg1})-(\ref{creatneg3}). The negativities are plotted in Fig.\ \ref{vergleich} and we see that the entanglement is distributed equally between the sectors in the infinite acceleration limit.

 For state $\Psi^\pm_{FF}$ we see again that entanglement is created in some sectors and  we can  write $N_{\Psi^\pm_{FF}}$ schematically as
\begin{align}
N_{\Psi^\pm_{FF}}=&N(1^+,0\,|\, 0,1^+)\nonumber\\
+&N(1^+1^-,0\,|\, 1^-,1^+)\nonumber\\
+&N(1^+,1^-\,|\, 0,1^+1^-)\nonumber\\
+&N(1^+1^-,1^-\,|\, 1^-,1^+1^-),
\end{align}
where
\begin{align}
&N(1^+,0\,|\, 0,1^+)= N(c^{\Psi_{FF}^+}_{\epsilon=1}),\\
&N(1^+1^-,0\,|\, 1^-,1^+)=\tan^2 (r_f^\omega) N(c^{\Psi_{FF}^+}_{\epsilon=1}),\label{creatneg51}\\
&N(1^+,1^-\,|\, 0,1^+1^-)=\tan^2 (r_f^\Omega) N(c^{\Psi_{FF}^+}_{\epsilon=1}),\\
&N(1^+1^-,1^-\,|\, 1^-,1^+1^-)=\tan^2 (r_f^\omega)\tan^2 (r_f^\Omega) N(c^{\Psi_{FF}^+}_{\epsilon=1})\label{creatneg53}
\end{align}
are the negativities of the different sectors. $N(1^+,0\,|\, 0,1^+)$ is the negativity of the sector, where the entanglement is initialized. When the acceleration increases from zero the negativity $N(1^+,0\,|\, 0,1^+)$ decreases while (\ref{creatneg51})-(\ref{creatneg53}) increase. In the infinite acceleration limit all sectors are equally entangled; see Fig.\ \ref{vergleich}. Comparing states $\Phi^\pm_{FF}$ and $\Psi^\pm_{FF}$, we note that the ``redistribution'' of entanglement is the same for small accelerations, but differs more and more with increasing acceleration, until finally different limiting values are approached. 

 We plot the entanglement in the  sectors that show creation of entanglement in Fig.\ \ref{vergleich}.  This effect is due to a symmetric production of particle-antiparticle pairs. We can, for example, write state $\Phi^\pm_{FF}$ schematically as $0,0\,|\,1^+,1^+$. Then, for nonzero acceleration, the Unruh effect, that in some sense can  be seen as pair production, populates states $0,1^-$ and $1^+,1^+1^-$ and thus creates entanglement [$N(0,1^-\,|\,1^+,1^+1^-)\neq 0$] in the sector $0,1^-\,|\,1^+,1^+1^-$. In the same manner there is entanglement created in the sectors $1^-,0\,|\,1^+1^-,1^+$ and $1^-,1^-\,|\,1^+1^-,1^+1^-$; see Fig.\ \ref{vergleich}. However, in general, not all states show this behavior. One state that does not show this feature is $\frac{1}{\sqrt{2}}\left(|1^F_\omega\rangle_U^+|1^F_\Omega\rangle_U^-+|1^F_\omega\rangle_U^-|1^F_\Omega\rangle_U^+\right)$. The reason why there is no entanglement generated is that for this state symmetric pair production would require a violation of the Pauli principle and therefore is forbidden.




\section{Boson-boson states}\label{appbb}
We calculate the negativitiy (\ref{negativity}) for the bosonic Bell states 
\begin{subequations}
\begin{align}
|\Psi^\pm_{BB}\rangle=&\frac{1}{\sqrt{2}}\left(|0_\omega\rangle_U|1_\Omega\rangle_U\pm|1_\omega\rangle_U|0_\Omega\rangle_U\right),\\
|\Phi^\pm_{BB}\rangle=&\frac{1}{\sqrt{2}}\left(|0_\omega\rangle_U|0_\Omega\rangle_U\pm|1_\omega\rangle_U|1_\Omega\rangle_U\right),
\end{align}
\end{subequations}
where $\omega$, $\Omega$ are the frequencies and $0$, $1$ the occupation numbers of the Unruh modes. The two modes $\omega$ and $\Omega$ undergo constant accelerations $a_\omega$ and $a_\Omega$. The acceleration parameters of the modes are denoted by $r^\omega$ and $r^\Omega$, respectively.

We start with state $\Phi^\pm_{BB}$ and for concreteness we consider $\Phi^+_{BB}$ that shows the same entanglement as $\Phi^\pm_{BB}$. So in the following we always denote $\Phi^+_{BB}$ by $\Phi^\pm_{BB}$ and similarly for the $\Psi^\pm$ states. The density matrix $\rho_{\Phi^\pm_{BB}}$ that we obtain after tracing out region $II$ is given by
\begin{align}\label{densitymatbb}
\rho_{\Phi^\pm_{BB}}=&\frac{1}{2} \sum_{m,n} a_m^2 a_n^2 |n m \rangle \langle n m| \nonumber\\
+&\frac{1}{2}\sum_{m,n} \bar{a}_m^2 \bar{a}_n^2 |(n+1) (m+1) \rangle \langle (n+1) (m+1)| \nonumber\\
+&\frac{1}{2} \sum_{m,n} a_m a_n\bar{a}_m \bar{a}_n  |n m \rangle \langle (n+1) (m+1)| +h.c.,
\end{align}
where $|n m \rangle=|n_\omega  \rangle_I\otimes|m_\Omega  \rangle_I$ and $a_n=a_n(r^\omega)=\tanh^{n} (r^\omega)\cosh^{-1} (r^\omega)$, $\bar{a}_n=\bar{a}_n(r^\omega) =\tanh^{n} (r^\omega)\cosh^{-2} (r^\omega) \sqrt{n+1}$. Accordingly, $a_m=a_{n=m}(r^\Omega)$ and $\bar{a}_m=\bar{a}_{n=m}(r^\Omega)$. The negativity can be obtained as a sum over the negativities for different values of $n, m$ ($N_{\Phi^\pm_{BB}}=\sum_{n,m}N_{\Phi^\pm_{BB}}^{(n,m)}$). This can be seen from the block diagonal structure of the partially transposed density matrix. The part of the partially transposed density matrix that contributes to $N_{\Phi^\pm_{BB}}^{(n,m)}$ and the negativity $N_{\Phi^\pm_{BB}}$ are given by
\begin{widetext}
\begin{equation}\label{relevant1bb}
\frac{1}{2} \frac{\tanh ^{2 n}(r^\omega) \tanh ^{2 m}(r^\Omega)}{\cosh^2 (r^\omega)\cosh^2 (r^\Omega)}\left(
\begin{array}{cc}
\left(1+\frac{(m+1) n }{\sinh ^{2}(r^\omega)\sinh ^{2}(r^\Omega)}\right)\tanh ^2(r^\Omega) & \frac{\sqrt{(m+1) (n+1)}}{\cosh (r^\omega)\cosh (r^\Omega)} \\
 \frac{\sqrt{(m+1) (n+1)}}{\cosh (r^\omega)\cosh (r^\Omega)}  & \left(1+\frac{(n+1) m }{\sinh ^{2}(r^\omega)\sinh ^{2}(r^\Omega)}\right)\tanh ^{2 }(r^\omega)  \\
\end{array}
\right)
\end{equation}
and
\begin{align}
N_{\Phi^\pm_{BB}}=&\sum_{n,m} N_{\Phi^\pm_{BB}}^{(n,m)}=\sum_{n} N_{\Phi^\pm_{BB}}^{(n,0)}+\sum_{m} N_{\Phi^\pm_{BB}}^{(0,m)}\nonumber\\
=&N_{\Phi^\pm_{BB}}^{(0)}+\sum_{n=1}^\infty\frac{\tanh ^{2 n}(r^\omega)}{4\cosh^2 (r^\omega)\cosh^2 ( r^\Omega)} ( \tanh ^2(r^\omega)+ \tanh ^2(r^\Omega)+ \frac{n}{\sinh^2 (r^\omega)\cosh^2 ( r^\Omega)} +\nonumber\\
 +&\frac{2}{\sinh^2 (r^\omega)\sinh^2 (r^\Omega)\cosh ( r^\omega)\cosh ( r^\Omega)}(\frac{n^2 }{4} \cosh ^2(r^\omega) \sinh ^2(r^\Omega) \tanh ^2(r^\Omega)+\sinh ^4(r^\omega) \sinh ^4(r^\Omega)\times\nonumber\\
 \times& (\frac{n}{2} +\frac{1}{4} \cosh ^2(r^\omega) \sinh ^2(r^\Omega) \tanh ^2(r^\Omega)+1)+\frac{n}{2} \sinh ^2(r^\omega) \cosh ^2(r^\omega) \sinh ^4(\text{r2}) \tanh ^2(r^\Omega)+\nonumber\\
 +&\sinh ^6(r^\omega) (\frac{1}{4} \tanh ^2(r^\omega) \sinh ^4(r^\Omega) \cosh ^2(r^\Omega)-\frac{1}{2} \sinh ^6(r^\Omega)))^\frac{1}{2})+\sum_{m=1}^\infty (\omega\leftrightarrow\Omega;\, n\rightarrow m),\label{bbgeneralnm}
\end{align}
 
\end{widetext}
where we used  that $N_{\Phi^\pm_{BB}}^{(n,m)}\neq 0$ only for either $n=0$ or $m=0$ or $n=m=0$. It can be seen that each of the $N_{\Phi^\pm_{BB}}^{(n,m)}$ is bounded from above by $N_{\Phi^\pm_{BB}}^{(0)}\equiv N_{\Phi^\pm_{BB}}^{(0,0)}$ that describes the entanglement between the modes we initially start with. For non-vanishing acceleration there is entanglement created between higher modes, i.e., $N_{\Phi^\pm_{BB}}^{(n,m)}\neq 0$, but this will be a small contribution compared to $N_{\Phi^\pm_{BB}}^{(0)}$; see Fig.\ \ref{piccreata}.

To obtain $N_{\Phi^\pm_{BB}}^{(0)}$, we set $n$ and $m$ to zero in (\ref{relevant1bb}).  The negativity $N_{\Phi^\pm_{BB}}^{(0)}$ can be written in the form 
\begin{equation}
N_{\Phi^\pm_{BB}}^{(0)}=2 N_{0,\omega}N_{0,\Omega}\, \gamma_{\Phi^\pm_{BB}}\left(n_B^\omega,\, n_B^\Omega\right),
\end{equation}
where $n_B^{\omega/\Omega}=(e^{\frac{\omega/\Omega}{T_{\omega/\Omega}}}-1)^{-1}$ is the Bose-Einstein distribution and $N_{0,\omega/\Omega}$ denotes the negativities if only mode $\omega/\Omega$ is accelerated. These  can be obtained from (\ref{bbgeneralnm}) by setting the acceleration parameter $r^{\Omega/\omega}$ to zero,
\begin{equation}\label{neg0bos1}
N_{\Phi^\pm_{BB}}^{(0)}(r^\omega,r^\Omega=0)\equiv N_{0,\omega}=\frac{1}{2}\frac{1}{\left(Z_B^{\omega}\right)^2},
\end{equation}
where  $Z_B^{\omega/\Omega}$ is the bosonic partition function (\ref{partbos}). Further, $\gamma_{\Phi^\pm_{BB}}$ is given by
\begin{equation}
\gamma_{\Phi^\pm_{BB}}=1-n_B^\omega  n_B^\Omega.
\end{equation}

Now we move to state $\Psi^\pm_{BB}$, where the relevant part of the partially transposed reduced density matrix is given by the following expression:
\begin{widetext}
\begin{equation}
\frac{1}{2} \frac{\tanh ^{2 n}(r^\omega)\tanh ^{2 m}(r^\Omega) }{\cosh^{2} (r^\omega)\cosh^{2} (r^\Omega)}\left(
\begin{array}{cc}
 \frac{m }{\sinh^{2} (r^\Omega)}+\frac{n }{\sinh^{2} (r^\omega)} & \frac{\sqrt{(m+1) (n+1)}}{\cosh (r^\omega)\cosh (r^\Omega)} \\
 \frac{\sqrt{(m+1) (n+1)}}{\cosh (r^\omega)\cosh (r^\Omega)}  & (m+1)\frac{\tanh ^{2 }(r^\omega) }{\cosh^{2} (r^\Omega)}+(n+1)\frac{\tanh ^{2 }(r^\Omega) }{\cosh^{2} (r^\omega)} \\
\end{array}
\right).
\end{equation}\begin{figure}
        \centering
        \begin{subfigure}[b]{0.47\columnwidth}
            \centering
            \includegraphics[width=\columnwidth]{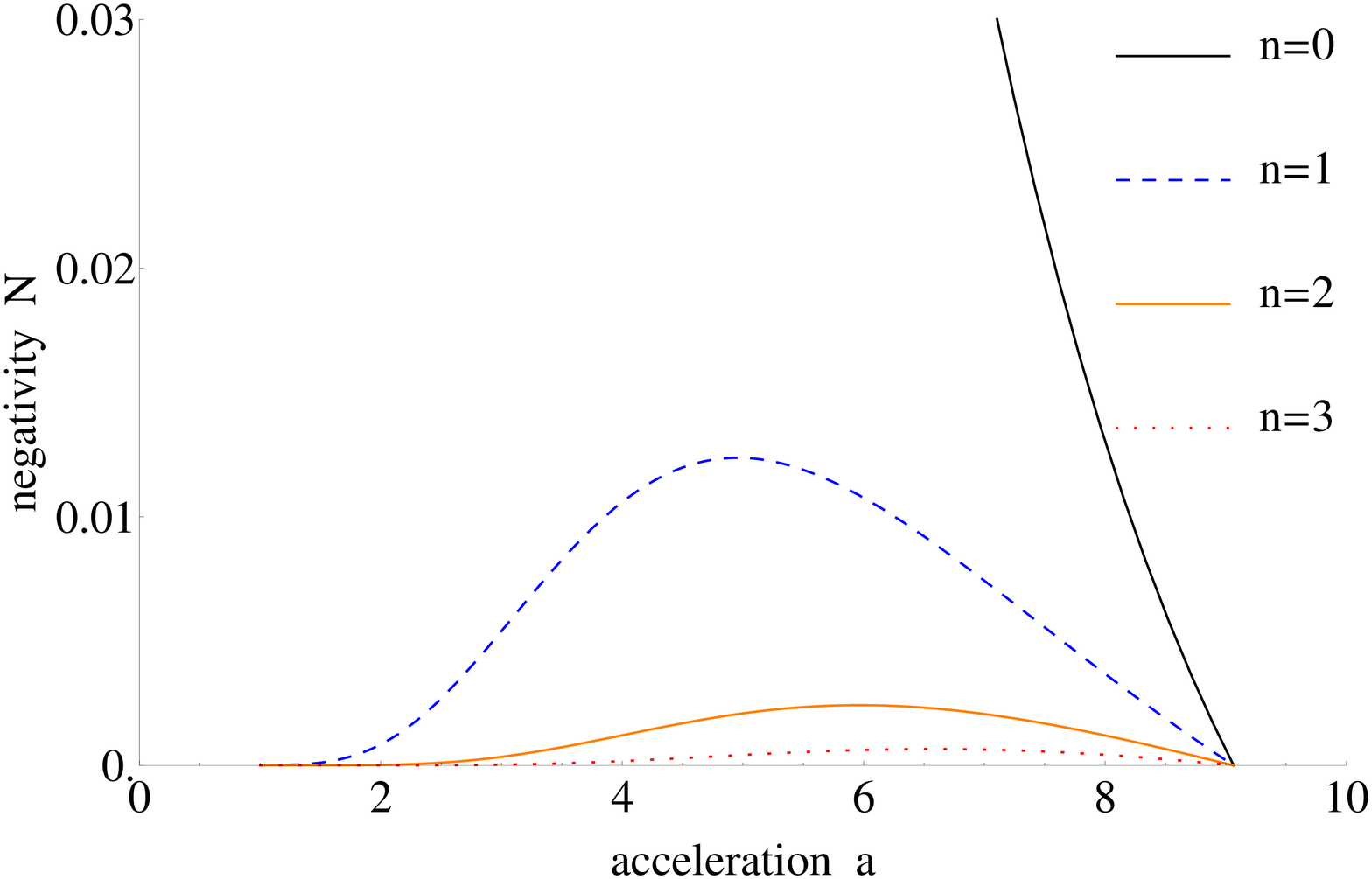}
            \caption[Network2]%
            {Negativities $N_{\Phi^\pm_{BB}}^{(n,m=0)}$. }\label{piccreata}    
        \end{subfigure}
        \hspace{1mm}
        \begin{subfigure}[b]{0.47\columnwidth}  
            \centering 
            \includegraphics[width=\columnwidth]{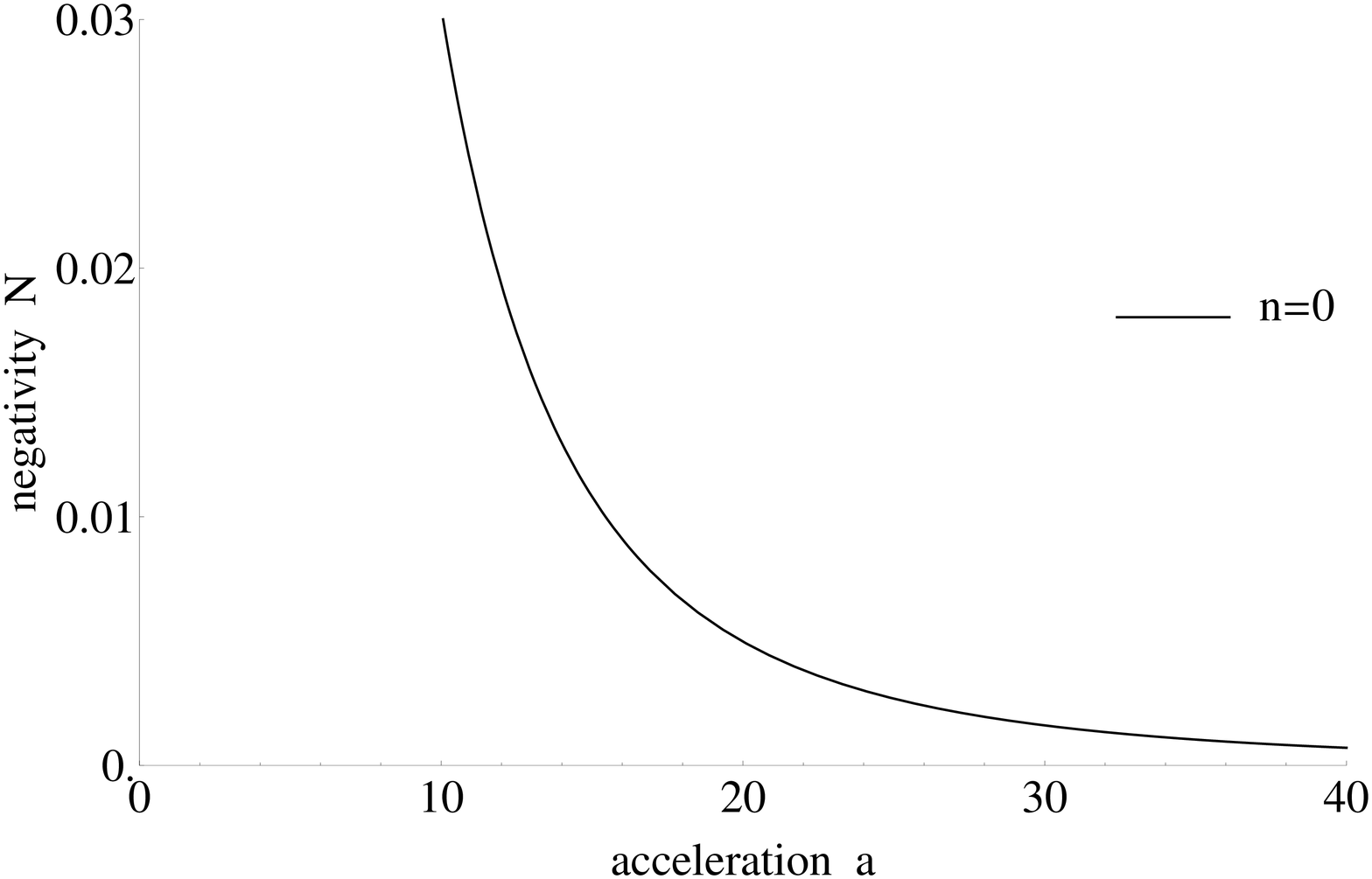}
            \caption[]%
            {Negativities $N_{\Psi^\pm_{BB}}^{(n,m=0)}$}\label{piccreatb}    
        \end{subfigure}
        \newline
        
        \vspace{0mm}
        
        \begin{subfigure}[b]{0.47\columnwidth}   
            \centering 
            \includegraphics[width=\columnwidth]{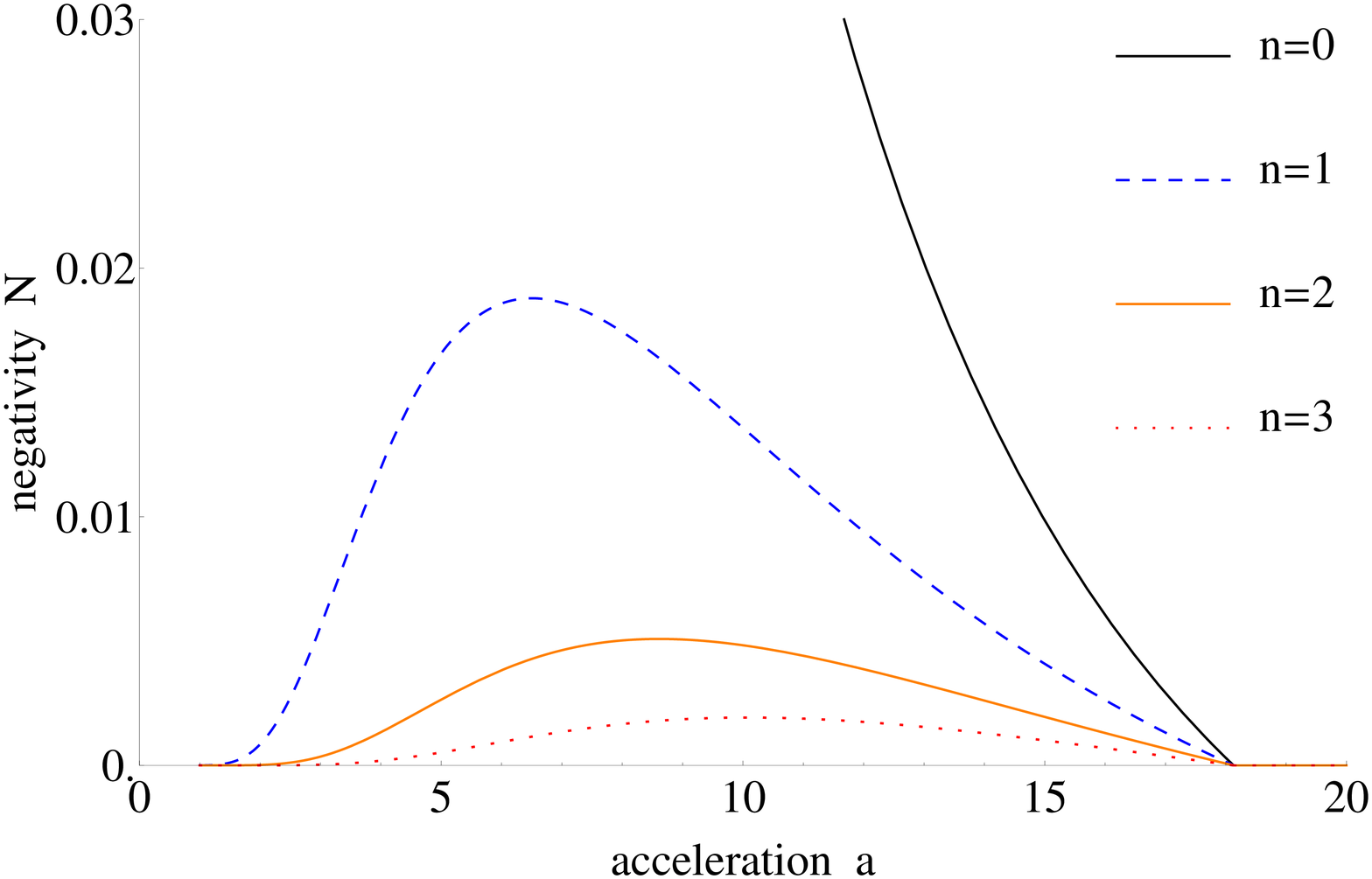}
            \caption[]%
            {Negativities $N_{\Phi^\pm_{BF}}^{(n)}$}\label{piccreatc}    
        \end{subfigure}
        \hspace{2mm}
        \begin{subfigure}[b]{0.47\columnwidth}   
            \centering 
            \includegraphics[width=\columnwidth]{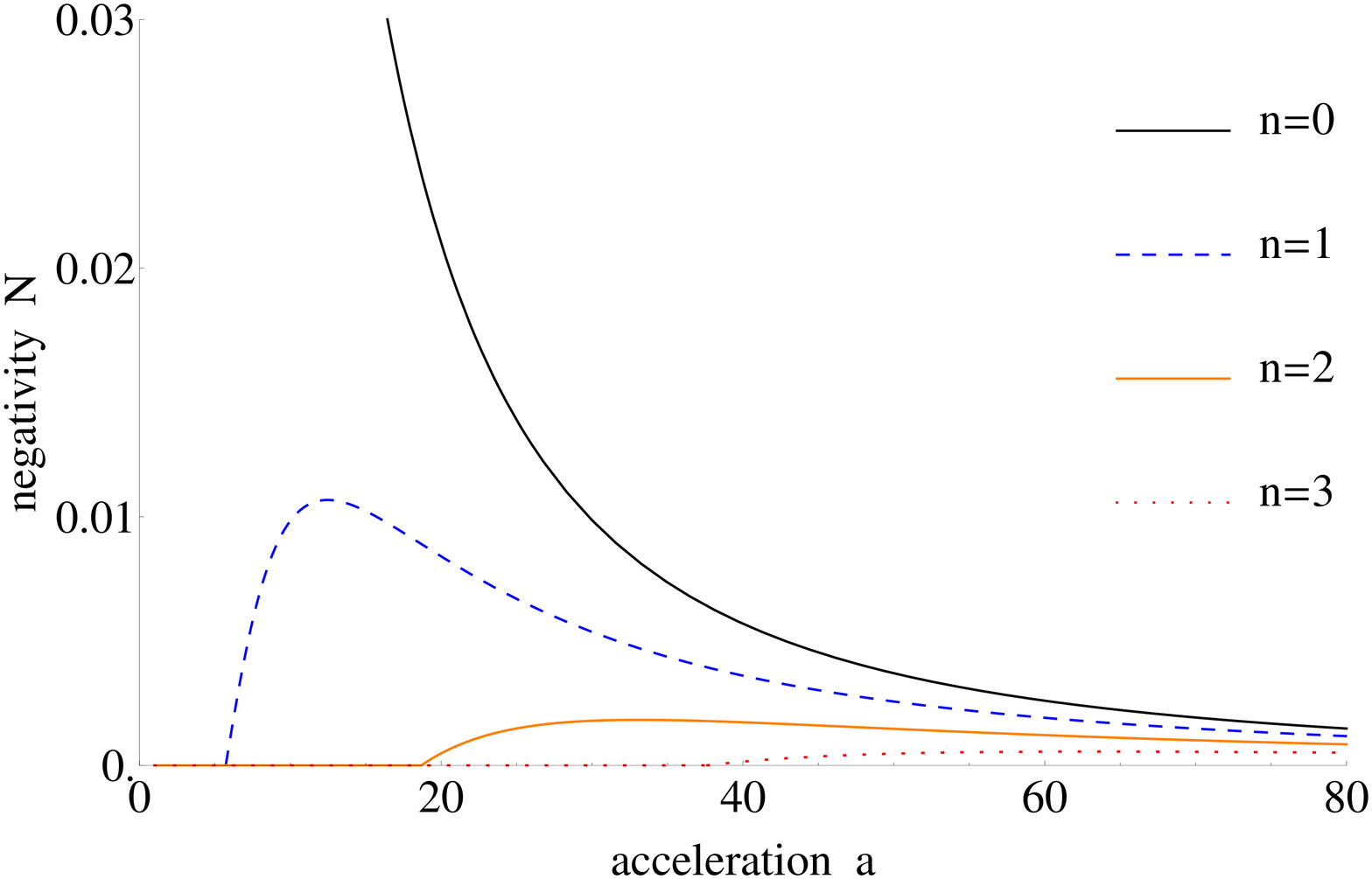}
            \caption[]%
            {Negativities $N_{\Psi^\pm_{BF}}^{(n)}$}\label{piccreatd}    
        \end{subfigure}
        \caption
        {(Color online) Negativities, where both observers are accelerated, plotted  against the acceleration  $a=a_\omega=a_\Omega$, measured in units of  $\frac{1}{L}$ (for some length scale $L$), for frequencies $\omega=\Omega=\frac{1}{L}$. While for states $\Phi^\pm_{BB}$, $\Phi^\pm_{BF}$, and $\Psi^\pm_{BF}$ there is entanglement created in initially nonentangled sectors (a), (c), and (d), there is no entanglement production (with this special choice of the acceleration parameter $r=r^\omega=r^\Omega$) for $\Psi^\pm_{BB}$ (b). Therefore in the generic case there is entanglement generated  in initially nonentangled sectors.} 
        \label{piccreat}
    \end{figure}

\end{widetext}
Contrary to (\ref{relevant1bb}), only the eigenvalues of the block $(n=0, m=0)$ can be negative [Fig.\ \ref{piccreatb}] for $r^\omega=r^\Omega$. In this case, the sum of the partial negativities $N_{\Psi^\pm_{BB}}=\sum_{n, m} N_{\Psi^\pm_{BB}}^{(n,m)}$ collapses to $N_{\Psi^\pm_{BB}}=N_{\Psi^\pm_{BB}}^{(0)}$ and we find
\begin{align}\label{n0psibb}
&N_{\Psi^\pm_{BB}}^{(0)}=\frac{1}{2}\frac{1}{\left(Z_B^\omega\right)^2}\frac{1}{\left(Z_B^\Omega\right)^2}\times\nonumber\\
&\times\left(\sqrt{Z_B^\omega Z_B^\Omega+\frac{1}{4}\left(n_B^\omega+n_B^\Omega\right)^2}-\frac{1}{2}\left(n_B^\omega+n_B^\Omega\right)\right).
\end{align}
In the case $r^\omega\neq r^\Omega$, we assume without loss of generality $r^\omega>r^\Omega$. Then the blocks for $m=0$ admit negative eigenvalues and we find the negativity
\begin{widetext}
\begin{align}
N_{\Psi^\pm_{BB}}=&N_{\Psi^\pm_{BB}}^{(0)}+\sum_{n=1}^\infty \frac{\tanh ^{2 n}(r^\omega)}{4\cosh^4 (r^\omega)\cosh^4 ( r^\Omega)}   (-n \coth ^2(r^\omega) \cosh ^2(r^\Omega)-(n+1) \sinh ^2(r^\Omega)-\sinh ^2(r^\omega)+\nonumber\\
+& \frac{1}{\sinh^2(r^\omega)} (\left(\frac{n}{2} \cosh (2 r^\omega) \cosh (2 r^\Omega)+\frac{n}{2}+ \sinh ^2(r^\omega) \sinh ^2(r^\Omega)+ \sinh ^4(r^\omega)\right)^2+\nonumber\\
-&\frac{1}{4} \sinh ^2(2 r^\omega) \sinh ^2(2 r^\Omega) \left(n (n+1) \sinh ^4(r^\Omega)-\sinh ^2(r^\omega) \sinh ^2(r^\Omega)\right))^{\frac{1}{2}}),\label{bbpsineg}
\end{align}
\end{widetext}
where $N_{\Psi^\pm_{BB}}^{(0)}$ is given by (\ref{n0psibb}).

\section{Boson-fermion states}\label{appbf}
We calculate the negativities for maximally entangled states of a bosonic mode entangled with a fermionic one. To start, we consider the states
\begin{subequations}
\begin{align}
|X_1\rangle=&\frac{1}{\sqrt{2}}\left(|0_\omega\rangle_U|1^F_\Omega\rangle_U^+ +|1_\omega\rangle_U|1^F_\Omega\rangle_U^-\right),\\
|X_2\rangle=&\frac{1}{\sqrt{2}}\left(|1_\omega\rangle_U^+|1^F_\Omega\rangle_U^-+|1_\omega\rangle_U^-|1^F_\Omega\rangle_U^+\right),
\end{align}
\end{subequations} 
where $F$ labels the fermionic mode, $\omega$, $\Omega$ are the frequencies and $0$, $1$ the occupation numbers of the Unruh modes. $+$ and $-$ refer to particles and antiparticles, respectively. The mode of frequency $\omega$ is bosonic while the mode of frequency $\Omega$ is fermionic. The respective acceleration parameters are given by $r=\text{arctanh} (e^{-\frac{\pi \omega}{a_\omega}})$ for the bosonic and $r_f=\arctan (e^{-\frac{\pi \Omega}{a_\Omega}})$ for the fermionic mode. As in Sec.\ \ref{appneg}, we carefully take into account the operator ordering for fermions, especially when we are performing partial traces. The relevant part of the partially transposed  reduced density matrices for these states can be computed to be
\begin{equation}
\frac{1}{2}\cos ^2(r_f)\frac{\tanh ^{2 n}(r) }{\cosh^2(r)}\left(
\begin{array}{cc}
   \frac{n}{\sinh^2(r)} & \sqrt{\frac{n+1}{\cosh^2(r)}} \\
  \sqrt{\frac{n+1}{\cosh^2(r)}}  &  \tanh ^{2}(r) \\
\end{array}
\right)
\end{equation}
for state $X_1$ and
\begin{align}
&\frac{1}{2}\cos ^2(r_f)\frac{\tanh ^{2 m+2 n}(r)}{\cosh^6(r)} \times\nonumber\\
&\times  \left(
\begin{array}{cc}
  n  &  \sqrt{(m+1) (n+1)}\\
  \sqrt{(m+1) (n+1)}  &  m  \\
\end{array}
\right)
\end{align}
for state $X_2$. We observe that the fermionic and the bosonic part factorize and so we find that the resulting negativity can be expressed in terms of negativities obtained from the cases of one accelerated observer. That is,
\begin{align}
N_{X_1}=& 2 N_{f} N_{b, 1},\\
N_{X_2}=& 2 N_{f} N_{b, 2},
\end{align}
where $N_{f}$ is the negativity $N_{f}=\frac{1}{2}\cos ^2(r_f) =\frac{1}{2}(Z_F^{\Omega})^{-1}$ and $N_{b, 1}$, $N_{b, 2}$ are given by
\begin{align}
N_{b, 1}=&\frac{1}{2}\frac{1}{\left(Z_B^\omega\right)^2}+\sum_{n=1}^\infty N_n,\label{Nn}\\
N_{b, 2}=&\frac{1}{\left(Z_B^\omega\right)^3}\sum_{n,m=0}^\infty e^{-\frac{2 \pi (n+m) }{a_\omega}}=\frac{1}{2}\frac{1}{Z_B^\omega},
\end{align}
i.e., the ones we obtain when we only accelerated the bosons. The $N_n$ in (\ref{Nn}) can be obtained by setting $r^\Omega=0$ in (\ref{relevant1bb}) and are given by
\begin{align}
N_n=&\frac{\tanh ^{2 n}(r)}{2 \cosh^2(r) }  ( \frac{n}{2\sinh^2(r)}+\frac{1}{2} \tanh ^2(r)+\nonumber\\
+&\sqrt{ \frac{n^2}{4\sinh^4(r)}+ \frac{\frac{n}{2}+\frac{1}{4}\sinh ^2(r) \tanh ^2(r)+1}{ \cosh^2(r) } }).
\end{align}

Next we move to  Bell states $\Psi^\pm_{BF}$ and $\Phi^\pm_{BF}$ that are given by
\begin{subequations}
\begin{align}
|\Psi^\pm_{BF}\rangle=&\frac{1}{\sqrt{2}}\left(|1_\omega\rangle_U|0^F_\Omega\rangle_U\pm|0_\omega\rangle_U|1^F_\Omega\rangle_U^+\right),\\
|\Phi^\pm_{BF}\rangle=&\frac{1}{\sqrt{2}}\left(|0_\omega\rangle_U|0^F_\Omega\rangle_U\pm|1_\omega\rangle_U|1^F_\Omega\rangle_U^+\right).
\end{align}
\end{subequations} 
The negativity of state $\Phi^\pm_{BF}$ is calculated in the following. For concreteness we carry the calculations out for state $\Phi^+_{BF}$. After obtaining the reduced density matrix $\rho_{\Phi^+_{BF}}$ by tracing out region $II$, 
\begin{widetext}
\begin{align}\label{densitymatbf}
\rho_{\Phi^+_{BF}}=\frac{\cos^2 (r_f)}{2} \lbrace&\sum_{n} \cos^2 (r_f) a_n^2  | n \rangle \langle  n|\otimes (| 0 0 \rangle \langle  0 0|+\tan^2 (r_f)| 1 0 \rangle \langle  1 0|) +\sum_{n}  \bar{a}_n^2 |(n+1) \rangle \langle (n+1) |\otimes | 1 0 \rangle \langle  1 0|\nonumber \\
+& \sum_{n} \cos (r_f)  a_n \bar{a}_n  |n  \rangle \langle (n+1) | \otimes | 0 0 \rangle \langle  1 0|+ h.c._{non diag.}\rbrace \nonumber \\
+\frac{\sin^2 (r_f)}{2}\lbrace&\sum_{n} \cos^2 (r_f) a_n^2  | n \rangle \langle  n|\otimes (| 0 1 \rangle \langle  0 1|+\tan^2 (r_f)| 1 1 \rangle \langle  1 1|) \nonumber +\sum_{n}  \bar{a}_n^2 |(n+1) \rangle \langle (n+1) |\otimes | 1 1 \rangle \langle  1 1| +\\
-& \sum_{n} \cos (r_f)  a_n \bar{a}_n  |n  \rangle \langle (n+1) | \otimes | 0 1 \rangle \langle  1 1|+ h.c._{non diag.}\rbrace,
\end{align}
\end{widetext}
where $a_n=a_n(r)=\tanh^{n} (r)\cosh^{-1} (r)$, $\bar{a}_n=\bar{a}_n(r) =\tanh^{n} (r)\cosh^{-2} (r) \sqrt{n+1}$, and the notation $| ij \rangle=| i_\Omega \rangle_I^+\otimes | j_\Omega \rangle_I^- $. Then after partial transposition, the relevant part of the reduced partially transposed density matrix  is of the form
\begin{equation}\label{densitybf1}
 \left(
\begin{array}{cc}
 c^{\Phi_{BF}^+}_{\epsilon=1} &  0\\
  0 &   \tan^2 (r_f)  c^{\Phi_{BF}^+}_{\epsilon=-1} \\
\end{array}
\right),
\end{equation}
where 
\begin{align}\label{densitybf12}
& c^{\Phi_{BF}^+}_{\epsilon}=\frac{1}{2}  \cos ^2(r_f) \frac{\tanh ^{2 n}(r) }{\cosh^2(r)}\times \nonumber\\
 &\times\left(
\begin{array}{cc}
 \cos ^2(r_f) \tanh ^{2}(r) &  \epsilon \cos (r_f)  \sqrt{\frac{n+1}{\cosh^2(r)}} \\
\epsilon \cos (r_f)  \sqrt{\frac{n+1}{\cosh^2(r)}}  &   \frac{n}{\sinh^2(r)}+\sin ^2(r_f) \\
\end{array}
\right).
\end{align}
Already at this stage we see that the fermionic contribution does not simply ``factor out'' as it was the case for states $X_1$ and $X_2$. Due to the block diagonal form of (\ref{densitybf1}), the negativity $N_{\Phi^+_{BF}}$ that equals $N_{\Phi^\pm_{BF}}$  can be written in the form
\begin{equation}\label{formofneg}
N_{\Phi^\pm_{BF}}=\sum_{n} N_{\Phi^\pm_{BF}}^{(n)}= (1+\tan^2 (r_f) )\sum_n \tilde{N}_{\Phi^\pm_{BF}}^{(n)},
\end{equation}
where from now on we identify $N_{\Phi^+_{BF}}$ and  $N_{\Phi^\pm_{BF}}$, and $\tilde{N}_{\Phi^\pm_{BF}}^{(n)}$ is the negativity calculated from (\ref{densitybf12}). Again, one can see that each $N_{\Phi^\pm_{BF}}^{(n)}$ is bounded from above by $N_{\Phi^\pm_{BF}}^{(0)}$; see Fig.\ \ref{piccreatc}. To obtain $N_{\Phi^\pm_{BF}}^{(0)}$ we have to calculate $\tilde{N}_{\Phi^\pm_{BF}}^{(0)}$
\begin{equation}
\tilde{N}_{\Phi^\pm_{BF}}^{(0)}= 2 N_{f}N_{0, \omega}  e^{\frac{\Omega}{T_\Omega}}  \left(\sqrt{n_F^\Omega n_B^\omega}-n_F^\Omega n_B^\omega\right),
\end{equation}
where $n_B^\omega=(e^{\frac{\omega }{T_\omega}}-1)^{-1}$ is the Bose-Einstein distribution, $N_{0, \omega}$ is given by (\ref{neg0bos1}), and $n_F^\Omega=(e^{\frac{\Omega }{T_\Omega}}+1)^{-1}$ is the Fermi-Dirac distribution and the $T_{\omega/\Omega}$ are the Unruh temperatures. Now we can use $N_{\Phi^\pm_{BF}}^{(0)}=(1+\tan^2 (r_f) )\tilde{N}_{\Phi^\pm_{BF}}^{(0)}$ to calculate
\begin{equation}
N_{\Phi^\pm_{BF}}^{(0)}=2 N_{f} N_{0, \omega}   \gamma_{\Phi^\pm_{BF}}(n_B^\omega,n_F^\Omega),
\end{equation}
where
\begin{equation}\label{gammaappphibf}
\gamma_{\Phi^\pm_{BF}}(n_B^\omega,n_F^\Omega)=\sqrt{\frac{n_B^\omega}{n_F^\Omega}}-n_B^\omega.
\end{equation}
The further $N_{\Phi^\pm_{BF}}^{(n)}$ for $n\neq 0$ can be obtained analytically and the negativity $N_{\Phi^\pm_{BF}}$ is obtained to be
\begin{widetext}
\begin{align}
N_{\Phi^\pm_{BF}}&=N_{\Phi^\pm_{BF}}^{(0)}+\sum_{n=1}^\infty\frac{ \tanh ^{2 n-2}(r) }{4\cosh^4(r)} (n+\sinh ^2(r) \cos ^2(r_f) \left(\tanh ^2(r)+\tan ^2(r_f)\right)+\nonumber\\
-&\sqrt{n^2+2 \sinh ^2(r) \left((n+2) \tanh ^2(r) \cos ^2(r_f)+n \sin ^2(r_f)\right)+\sinh ^4(r) \left(\sin ^2(r_f)-\tanh ^2(r) \cos ^2(r_f)\right)^2}).
\end{align}
\end{widetext} 
 To obtain a condition for vanishing negativity, we have a look at  (\ref{gammaappphibf}) and realize that the condition for entanglement can be written as
\begin{equation}
n_B^\omega n_F^\Omega\leq 1.
\end{equation}

Finally, we calculate the negativity of state $\Psi^\pm_{BF}$, where we again, for the sake of concreteness, consider $\Psi^+_{BF}$. The relevant part of the reduced partially transposed density matrix is of the form
\begin{equation}\label{relevant4}
\left(
\begin{array}{cc}
c^{\Psi_{BF}^+}_{\epsilon=1} &  0\\
  0 &   \tan^2 (r_f) c^{\Psi_{BF}^+}_{\epsilon=-1} \\
\end{array}
\right),
\end{equation}
where
\begin{align}\label{relevant4rho}
 &c^{\Psi_{BF}^+}_{\epsilon}=\frac{1}{2}  \cos ^2(r_f) \frac{\tanh ^{2 n}(r)}{\cosh^2(r)}\times\nonumber\\
 &\times\left(
\begin{array}{cc}
 \cos ^2(r_f) \frac{n}{\sinh^2(r)} & \epsilon \cos (r_f)  \sqrt{\frac{n+1}{\cosh^2(r)}}\\
\epsilon \cos (r_f)  \sqrt{\frac{n+1}{\cosh^2(r)}}  &   \frac{n+1}{\cosh^2(r)}\sin ^2(r_f) + \tanh ^{2}(r) \\
\end{array}
\right).
\end{align}
Due to the structure of (\ref{relevant4}) the negativity again has the form (\ref{formofneg}). Then $N_{\Psi^\pm_{BF}}$ is calculated to be
\begin{widetext}
\begin{align}
N_{\Psi^\pm_{BF}}=&N_{\Psi^\pm_{BF}}^{(0)}+\sum_{n=1}^\infty\frac{ \tanh ^{2 n-2}(r) }{4\cosh^4(r)} (\tanh ^2(r) \left((n+1) \sin ^2(r_f)+\sinh ^2(r)\right)+n \cos ^2(r_f)+\nonumber\\
-&2(\frac{n^2}{4} \cos ^4(r_f)+\tanh ^4(r) (\frac{n+1}{2} \sinh ^2(r) \sin ^2(r_f)+\frac{1}{4} ( n+1)^2 \sin ^4(r_f)+\frac{1}{4} \sinh ^4(r))+\nonumber\\
+&\tanh ^2(r) \cos ^2(r_f) ((\frac{n	}{2}+1) \sinh ^2(r)-\frac{n+1}{2} n \sin ^2(r_f)))^\frac{1}{2}),
\end{align}
\end{widetext}
where
\begin{equation}
N_{\Psi^\pm_{BF}}^{(0)}=\frac{1}{2}\frac{1}{Z_F^\Omega}\frac{1}{\left( Z_B^\omega \right)^2}.
\end{equation}
$N_{\Psi^\pm_{BF}}^{(0)}$ again gives an upper bound on all the $N_{\Psi^\pm_{BF}}^{(n)}$ and an lower bound on $N_{\Psi^\pm_{BF}}$;  see Fig.\ \ref{piccreatd}.

\section{Near horizon limit for a Schwarzschild black hole}\label{appbh}

In the presence of a  Schwarzschild black hole the spacetime outside the black hole is characterized by the Schwarzschild metric,
\begin{equation}\label{schwarz}
ds^2=\left(1-\frac{R_S}{r}\right)dt^2-\frac{1}{1-\frac{R_S}{r}}dr^2-r^2d\Omega^2,
\end{equation}
where $G$ is the gravitational constant, $M$ is the mass of the black hole, $R_S=2GM$ is the Schwarzschild radius and $d\Omega^2$ is the line element of the unit 2-sphere. In order to obtain the limiting form of (\ref{schwarz}) close to the horizon of a Schwarzschild black hole, we consider an observer placed at $r=r_0$ with proper time $\eta=\left(1-2GM/r_0\right)t$. Introducing
\begin{equation}
\rho^2=8GM\left(r-2GM\right),
\end{equation}
we obtain  the following metric up to terms of order $\frac{1}{2GM}$ in the near horizon limit
\begin{equation}\label{nearhorizon}
ds^2=ds_{R}^2-ds_2^2,
\end{equation}
where
\begin{align}
ds_{R}^2=& \frac{1}{16 G^2M^2} \left(1-\frac{2GM}{r_0}\right)^{-1}\rho^2d\eta^2-d\rho^2 ,\\
ds_2^2=&(2GM)^2d\Omega^2.
\end{align}
So (\ref{schwarz}) reduces to the product of two-dimensional Rindler space ($ds_{R}^2$) and a 2-sphere of radius $2GM$ ($ds_{2}^2$). Comparing $ds_{R}^2$ to the two-dimensional Rindler metric (\ref{metricrindler}), we see that the acceleration $a$ experienced by an observer at fixed position $r=r_0$ is given by
\begin{equation}\label{bhacc}
a=\frac{1}{4GM}\left(1-\frac{2GM}{r_0}\right)^{-\frac{1}{2}}.
\end{equation}

 To extend the considerations of Secs.\ \ref{ffe}, \ref{bbe}, and \ref{bfe} in 2d Rindler space to this spacetime, we consider the wave equation for a massless scalar field $\psi$ that is given by $\Box \psi=0$. In the near horizon limit (\ref{nearhorizon}) we can write
\begin{equation}
(\Box_{R}-\Delta_{S^2}) \psi(\eta,\rho,\phi,\theta)=0,
\end{equation}
where $\phi$, $\theta$ are angular coordinates, $\Box_{R}$ is the d'Alembertian of 2d Rindler space, and $\Delta_{S^2}$ is the Laplacian of the 2-sphere. We are looking for solutions of the form 
\begin{equation}\label{statebh}
 \psi(\eta,\rho,\phi,\theta)=\psi_{rad}(\eta,\rho)\psi_{ang}(\phi,\theta),
\end{equation}
that satisfy
\begin{align}
\Box_{R}\psi_{rad}(\eta,\rho)=&0, \label{rindlermodes} \\
\Delta_{S^2}\psi_{ang}(\phi,\theta)=&0. \label{angularmodes}
\end{align}
The solutions of (\ref{rindlermodes}) are the well known solutions of the Klein Gordon equation in Rindler space that we used above. The eigenfunctions of $\Delta_{S^2}$ are given by the spherical harmonics $Y^m_l(\phi, \theta)$. The eigenvalues are $l(l+1)$. So we pick the eigenfunctions with $l=0$, i.e., 
\begin{equation}
\psi_{ang}(\phi,\theta)=e^{im\phi}P_{l=0}^m(\cos(\theta))=1,
\end{equation}
where the $P^m_l(\cos(\theta))$ are the associated Legendre polynomials. We conclude that for the choice $l=0$, i.e., zero angular momentum, we can describe the near horizon limit by restricting our considerations to  2d Rindler space. 

Therefore, in the following we restrict ourselves to wave functions $\psi$ of vanishing angular momentum satisfying (\ref{statebh}). When we consider maximally entangled fermion states (\ref{statesfermi}) to hover over a black hole at some distance $d=r_0-R_S$ from the horizon, the system can be described in 2d Rindler space (for some more details on this correspondence, see \cite{martin2010unveiling}). The analog of the Rindler vacuum $|0\rangle_I$ is the Boulware vacuum $|0\rangle_B$ and the Unruh vacuum $|0\rangle_U$ corresponds to the Hartle-Hawking vacuum $|0\rangle_H$. Further, the physical effect that causes the degradation of entanglement is now the Hawking effect.   Near a black hole of mass $M$ the acceleration $a$ in (\ref{metricrindler}) is set by (\ref{bhacc}). So, we see that the limit of infinite acceleration corresponds to the limit of $r_0$ approaching $R_S$.   Considering an observer stationary at a radial distance of $r_0$, one can write the acceleration parameter as 
\begin{align}
r=&\text{arctanh} (e^{- \frac{\omega_g}{2}\sqrt{1-\frac{R_S}{r_0}}}),\label{accpara1}\\
r_f=&\arctan (e^{- \frac{\omega_g}{2}\sqrt{1-\frac{R_S}{r_0}}}),\label{accpara2}
\end{align}
where $\omega_g$ and $\Omega_g$ are the rescaled frequencies $\omega_g=4\pi R_S\omega$ and $\Omega_g=4\pi R_S\Omega$. Plugging the acceleration parameters (\ref{accpara1}) and (\ref{accpara2}) into the expressions for the negativities we obtained above, one obtains the negativities of the respective states in the case that the acceleration is due to the presence of a black hole.

\bibliography{literature_rqi}

\end{document}